\documentclass[10pt, twocolumn]{iopart}

\usepackage{comment}
\usepackage{xcolor}
\usepackage{array}
\usepackage{arydshln}
\expandafter\let\csname equation*\endcsname=\relax 
\expandafter\let\csname endequation*\endcsname=\relax 
\usepackage{amsmath}
\usepackage{amssymb}
\usepackage{subcaption}
\usepackage{gensymb}
\usepackage{cancel}
\usepackage[margin=0.6in]{geometry}
\usepackage{makecell}
\usepackage{graphicx,booktabs}
\usepackage{multicol}
\usepackage[utf8]{inputenc}
\usepackage[T1]{fontenc}

\usepackage[separate-uncertainty=true]{siunitx}
\DeclareSIUnit\torr{torr}

\begin{document}

\twocolumn[
  \begin{@twocolumnfalse}
    \title[Enhanced collisionality-driven pedestal transport]{Enhanced pedestal transport driven by edge collisionality on Alcator C-Mod and its role in regulating H-mode pedestal gradients}

\author{M.A. Miller$^1$,
        J.W. Hughes$^1$,
        A.M. Rosenthal$^2$,
        S. Mordijck$^3$,
        R. Reksoatmodjo$^4$,
        M. Wigram$^1$,
        J. Dunsmore$^1$,
        F. Sciortino$^5$,
        R.S. Wilcox$^6$,
        T. Odstrčil$^7$}

\address{$^1$MIT Plasma Science and Fusion Center, Cambridge, MA 02139, USA
}
\address{$^2$Commonwealth Fusion Systems, Devens, MA 01434, USA
}
\address{$^3$William \& Mary, Williamsburg, VA 23188, USA
}
\address{$^4$Lawrence Livermore National Laboratory, Livermore, CA 94550, USA
}
\address{$^5$Proxima Fusion, 81671 München, Germany
}
\address{$^6$Oak Ridge National Laboratory, Oak Ridge, TN 37831, USA
}
\address{$^7$General Atomics, San Diego, CA 92186, USA
}
\ead{millerma@mit.edu}

\maketitle
    \label{sec:abstract}

\begin{abstract}
Experimental measurements of plasma and neutral profiles across the pedestal are used in conjunction with 2D edge modeling to examine pedestal stiffness in Alcator C-Mod H-mode plasmas. Enhanced D$_\alpha$ (EDA) experiments on Alcator C-Mod observed pedestal degradation and loss in confinement below a critical value of net power crossing the separatrix, $P_\mathrm{net} = P_\mathrm{net}^\mathrm{crit} \approx 2.3$ MW, in the absence of any external fueling. New analysis of ionization and particle flux profiles reveal saturation of the pedestal electron density, $n_{e}^\mathrm{ped}$, despite continuous increases in ionization throughout the pedestal, inversely related to $P_\mathrm{net}$. A limit to the pedestal $\nabla n_{e}$ emerges as the particle flux, $\Gamma_{D}$, continues to grow, implying increases in the effective particle diffusivity, $D_\mathrm{eff}$. This is well-correlated with the separatrix collisionality, $\nu^{*}_\mathrm{sep}$ and a turbulence control parameter, $\alpha_{t}$, implying a possible transition in type of turbulence. The transition is well correlated with the experimentally observed value of $P_\mathrm{net}^\mathrm{crit}$. SOLPS-ITER modeling is performed for select discharges from the power scan, constrained with experimental electron and neutral densities, measured at the outer midplane. The modeling confirms general growth in $D_\mathrm{eff}$, consistent with experimental findings, and additionally suggests even larger growth in $\chi_{e}$ at the same $P_\mathrm{net}^\mathrm{crit}$. 
\end{abstract}

  \end{@twocolumnfalse}
]

\section{Introduction}
\label{sec:intro}

The physics that determine plasma profiles in the edge of tokamak plasmas remains among the most elusive in tokamak physics. It is known that at sufficient input power, a transition to a high-confinement mode (H-mode) occurs, resulting in a substantial reduction in edge transport, increasing pressure gradients, and forming the so-called ``pedestal" \cite{asdex_team_h-mode_1989}. It is well-known that achievable core pressures are very sensitive to the conditions at the boundary \cite{kotschenreuther_quantitative_1995, greenwald_h_1997, kinsey_iter_2011, frassinetti_global_2017, rodriguez-fernandez_predictions_2020}. Next-generation devices \cite{physics_chapter_2007, rodriguez-fernandez_overview_2022} seeking to maximize fusion gain rely on a robust edge pedestal. If the H-mode pedestal is to be exploited for high fusion gain, confidence in understanding the mechanisms determining its structure and the ability to form a desired type of pedestal must increase, especially in view of integrating this edge with a viable heat exhaust scrape-off layer (SOL) and divertor solution. 

Some models have had reasonable success in predicting the pressure pedestal structure of H-modes exhibiting Type-I edge localized modes (ELMs) in conventional aspect ratio machines \cite{snyder_pedestal_2009, groebner_limits_2010, walk_characterization_2012, snyder_high_2019}. Regardless, prediction of pedestal structure in turbulence-limited H-modes without ELMs is less certain \cite{viezzer_prospects_2023}. Density pedestals in particular are especially poorly understood, largely as a result of a lack of edge ionization source measurements. It is unclear to what extent neutral fueling, as opposed to particle transport, is responsible for the build up of edge pedestal density gradients \cite{hughes_advances_2006, mordijck_overview_2020}. Certain models suggest that the density pedestal width may be strongly linked to the neutral penetration depth \cite{mahdavi_physics_2003, groebner_role_2002}. While this may hold in some H-mode regimes, it cannot explain the width of all density pedestals \cite{hughes_advances_2006, hughes_edge_2007}. In ohmic L-modes, turbulent transport has been identified as important for regulating edge gradients \cite{labombard_critical_2008, labombard_evidence_2005}, and it is unknown to what degree turbulence is responsible for holding H-mode gradients at or near marginal stability. Since the very small length scales in the pedestal makes direct gyrokinetic simulation of these instabilities challenging, reduced-order models can serve as alternatives. Knowledge of the distribution of sources in the edge, in particular the ionization source, is an asset to testing reduced-order fueling and transport models.

It is also thought that understanding and predicting the density and temperature pedestals individually is crucial for differentiating between the different flavors of H-mode \cite{hughes_pedestal_2013, faitsch_analysis_2023}. The density at the edge in particular is a very important parameter for determining how particles and heat cross onto open field lines and affect the survivability of the divertor \cite{labombard_cross-field_2000, lackner_role_1993}. 

Alcator C-Mod operated until 2016, and it routinely achieved the highest plasma densities of current tokamaks, fairly close to and often exceeding densities proposed for reactors \cite{greenwald_20_2014, hughes_access_2018, mordijck_overview_2020}. As edge plasma density rises, opaqueness to neutrals increases, and fueling through gas puffing may become a challenge. As opaqueness increases, the ionization profile may move radially outward onto the unconfined plasma. If this is the case for future reactors, their ability to sustain large density gradients in the edge may be significantly reduced. A proposed metric for plasma opaqueness, $\eta = a \times n$, where $a$ is the plasma minor radius and $n$ is the plasma density, puts Alcator C-Mod's opaqueness close to that expected on SPARC and ITER \cite{mordijck_overview_2020, reksoatmodjo_role_2021}. Higher $a$ and $\overline{n}_{e}$ are thought to increase the difficulty with which an edge neutral can reach the core. Ionization profiles on Alcator C-Mod thus provide insight into how ionization could look in future devices. Furthermore, pedestal resiliency to fueling and evidence of critical edge gradients \cite{hughes_edge_2007, labombard_critical_2008} on C-Mod may be a feature of high-density devices with which future reactors need to contend. 

To this end, this work attempts to study the structure of the density pedestal on Alcator C-Mod using edge ionization and plasma density measurements. From inferences of particle transport and high-resolution Thomson scattering measurements of pedestal structure, this paper finds particle transport, as well as heat transport, in the pedestal to be largely influential in determining critical edge gradients and the achievable pedestal densities in a number of H-modes on Alcator C-Mod.

This paper begins by outlining the set of diagnostics used to create a database of edge ionization and pedestal profiles. Section \ref{sec:fueling} then compares experimental ionization measurements with density measurements from a particular set of experiments at fixed magnetic field, plasma current, and plasma shape to make comments about observed pedestal stiffness, and the accompanying increase in particle transport. Section \ref{sec:os} then considers the pedestal operational space, and makes claims about how the C-Mod plasma moves around this operational space. It identifies collisionality, particularly at the edge, as a key driver for determining both the density and temperature pedestal. To supplement these experimental observations, Section \ref{sec:solps} focuses on results from a set of simulations containing high-fidelity and experimentally-validated 2D calculations of the ionization source. The paper then ends with discussion on implications of these results for next-generation devices, and future steps that will be taken to develop a predictive model for next-generation density pedestals.

\section{Experimental ionization and particle flux profiles in a set of magnetic balance experiments}
\label{sec:experiment}

\begin{figure}[h!]
\centering
\includegraphics[width=0.6\columnwidth]{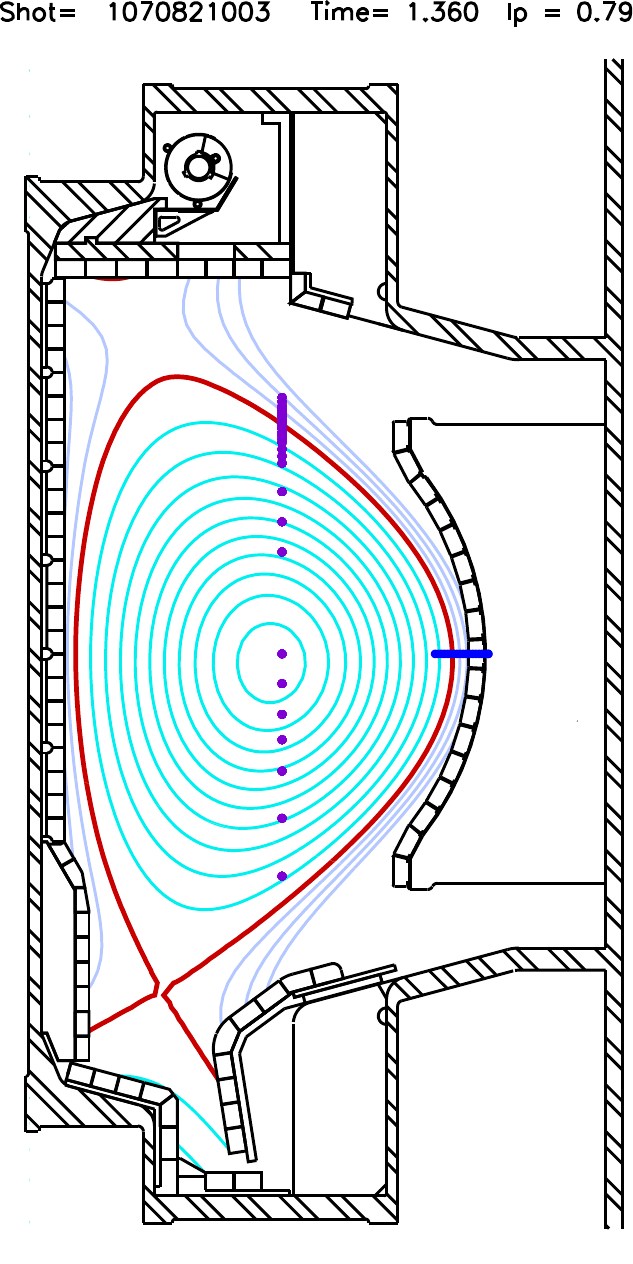}
\caption{Typical magnetic equilibrium for LSN discharges in effective power scan. The thick red line demarcates the last closed flux surface, separating the closed field lines, shown in cyan, and the open field lines, shown in dark blue. The TS system in purple views vertically down the plasma. The sparser core system views across the majority of the core plasma. The edge TS system (ETS) views the area around the separatrix, located poloidally near the crown of the plasma in the LSN configuration. The LYMID system, shown in bright blue, is located exactly at the midplane and views toroidally along 20 chords.}
\label{fig:xs}
\end{figure}

In 2007, a series of experiments were performed on Alcator C-Mod to study the effect of switching the magnetic geometry from lower single null (LSN) to upper single null (USN) on the EDA H-mode regime. The EDA H-mode is a type of H-mode not limited by peeling-ballooning MHD modes, as the more typical type-I ELMing H-modes are \cite{hubbard_pedestal_2001, hughes_pedestal_2013}. To study the effect of the magnetic drift on EDA H-mode pedestal quality, for a fixed toroidal magnetic field, $B_{t}$ = 5.4 T, plasma current, $I_{P}$ = 0.8 MA, and $\nabla B$-drift direction, the active null was moved from the lower chamber to the upper chamber, effectively modifying the outer midplane distance between the separatrices passing through the two X-points, $\Delta R_\mathrm{sep}$, from around -5\,mm to +5\,mm. The current work focuses solely on plasmas solidly in LSN, with $\Delta R_\mathrm{sep} \approx -5$ mm. These were all obtained at relatively fixed shape, with elongation, $\kappa \approx 1.6$, and triangularity, $\delta \approx 0.5$. They had a major radius, $R_{0} \approx 0.67$ m and a minor radius, $a \approx 0.22$ m. Figure \ref{fig:xs} shows a typical magnetic equilibrium for these plasmas, showing also views of the diagnostics to be described in Sections \ref{subsec:lymid} and \ref{subsec:ETS}. As was the case for many H-modes on Alcator C-Mod, gas was only injected in the L-mode phase, and the target density was sustained in the H-mode without any additional gas request on the controller. These plasmas also received no external impurity injection.

This paper presents a self-consistent analysis of particle and neutral pedestal transport from these experiments. It does so by combining radially-resolved edge electron density and temperature ($n_{e}$, $T_{e}$) measurements from Thomson Scattering (TS) with measurements of neutral emissivity from a Lyman-$\alpha$ camera, $\epsilon_\mathrm{Ly_{\alpha}}$, to infer neutral density, $n_{0}$, ionization source, $S_\mathrm{ion}$, radial particle flux, $\Gamma_{D}$, and effective diffusivity, $D_\mathrm{eff}$.

\subsection{Pedestal degradation with net power}
\label{subsec:magbal}

It was observed that these plasmas, particularly their pedestals, were highly sensitive to variation in the net power flowing through the scrape-off layer, $P_\mathrm{net}$, calculated using $P_\mathrm{net} = P_\mathrm{loss} - P_\mathrm{rad}$, where the loss power $P_\mathrm{loss} = P_\mathrm{oh} + P_\mathrm{aux} - dW/dt$ and $P_\mathrm{rad}$ is the power radiated in the core. Auxiliary power $P_\mathrm{aux}$ in this case is just the total injected ion cyclotron range of frequencies (ICRF) power, $P_\mathrm{RF}$. $P_\mathrm{oh}$ is the ohmic power, and $dW/dt$ is the time derivative of the stored energy, $W_\mathrm{MHD}$. 

\begin{figure*}[h!]
\centering
\includegraphics[width=2\columnwidth]{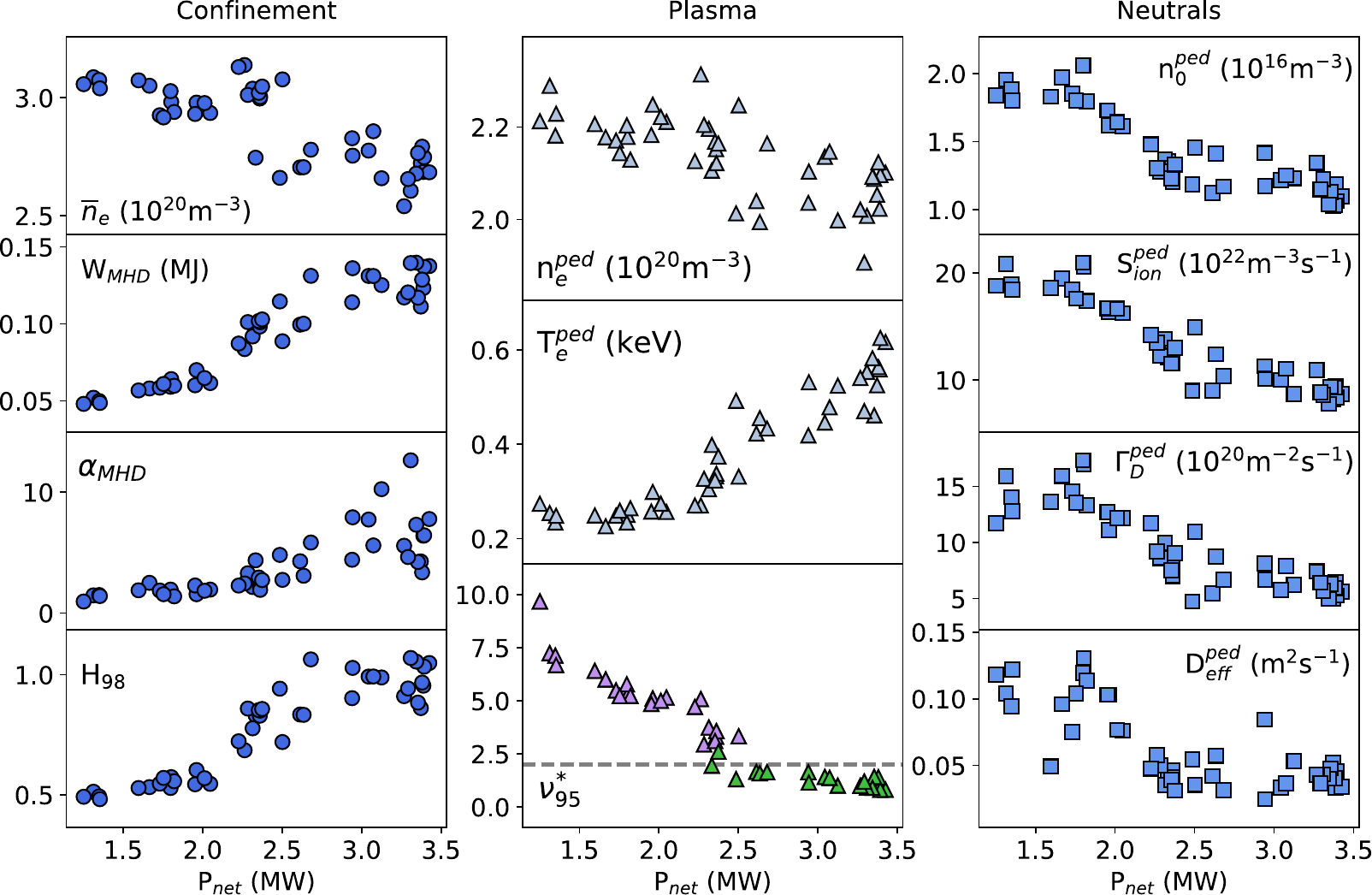}
\caption{Various experimentally-determined quantities plotted against $P_\mathrm{net}$. The plots in the left panel show quantities typically linked with plasma confinement in dark blue circles. The center panels show variation of plasma quantities at the pedestal top in gray triangles, with the exception of the bottom panel, showing collisionality at $\psi_{n} = 0.95$, for which a distinction in color is made. Purple triangles have $\nu^{*}_{95} > 2$, while green triangles have $\nu^{*}_{95} < 2$. This transition value roughly coincides with a critical value of $P_\mathrm{net} \approx$ 2.3 MW, at which all other quantities exhibit step-like changes. Finally, the panel on the right shows variation in quantities directly related to neutrals using light blue squares.}
\label{fig:pnet}
\end{figure*}

Figure \ref{fig:pnet} shows an effective scan in $P_\mathrm{net}$ from 1.3 MW to 3.4 MW. It shows large changes to many quantities in these discharges as $P_\mathrm{net}$ approaches and crosses a critical value of $P_\mathrm{net}$, $P_\mathrm{net}^\mathrm{crit} \approx$ 2.3 MW. The left panel of Figure \ref{fig:pnet} shows that the line-integrated density, $\overline{n}_{e}$, drops by $\sim$8\%, simultaneous with large increases in $W_\mathrm{MHD}$, the maximum normalized pressure gradient, $\alpha_\mathrm{MHD}$, and the normalized confinement factor, $H_\mathrm{98,y2}$ ($H_{98}$) \cite{iter_physics_1999}. Recalling the expression for the opaqueness heuristic introduced in Section \ref{sec:intro}, the plot of $\overline{n}_{e}$ also gives information about the opaqueness in these shots. Multiplying $\overline{n}_{e}$ by $a$ for these shots gives a range in opaqueness from $5.5 - 6.8 \times 10^{19}$m$^{-2}$. While these values are among the highest of current devices, they are not the highest achieved on C-Mod. For example, plasmas in the final campaigns of C-Mod achieved opaqueness values of up to $1.1 \times 10^{20}$m$^{-2}$ \cite{Hughes_2018}. For reference, recent plasmas with Greenwald fraction larger than unity on DIII-D achieved an opaqueness just under $4.4 \times 10^{19}$m$^{-2}$ \cite{ding_high-density_2024}. ITER and SPARC, on the other hand, are expected to operate at opaqueness values of $2.2 \times 10^{20}$ m$^{-2}$ and $1.7 \times 10^{20}$ m$^{-2}$, respectively \cite{campbell_iter_nodate, rodriguez-fernandez_predictions_2020}. If instead, the plasma density is taken as $n = \frac{1}{2}(n_{e}^\mathrm{sep} + n_{e}^\mathrm{ped})$, as suggested in \cite{mordijck_overview_2020}, the opaqueness for these discharges ranges from from $3.1 - 4.3 \times 10^{19}$m$^{-2}$ and that predicted for SPARC and ITER are $1.6 \times 10^{20}$ m$^{-2}$ and $1.5 \times 10^{20}$ m$^{-2}$, respectively \cite{hughes_projections_2020, mordijck_impact_2024}.

The center panel of Figure \ref{fig:pnet} shows changes in pedestal plasma parameters as measured by the ETS diagnostic, outlined in Section \ref{subsec:ETS}. There is a large drop in the collisionality at $\psi_{n}$ = 0.95, $\nu^{*}_{95}$, resulting from a very slight drop in $n_{e}^\mathrm{ped}$, but a large rise in $T_{e}^\mathrm{ped}$. Each of these pedestal quantities is computed at their respective pedestal top, defined by the fit function described in Section \ref{subsec:ETS} rather than at a particular flux surface. In general, these are at different locations for $n_{e}$ and $T_{e}$. The bottom plot of the center panel differentiates, with color, discharges that are at high values of $\nu^{*}_{95} > 2$ and low values of $\nu^{*}_{95} < 2$. This transition in collisionality occurs at around a value of $P_\mathrm{net} \approx$ 2.3 MW, at which point all the quantities in Figure \ref{fig:pnet} show clear changes.

Revisiting this particular experiment through inspection of edge ionization profiles reveals the role of plasma transport in regulating the pedestal and thus, the H-mode quality. The right panel of Figure \ref{fig:pnet} alludes to this, showing large drops in the inferred $n_{0}$, $S_\mathrm{ion}$, $\Gamma_{D}$, and $D_\mathrm{eff}$, all at the top of the pedestal. The discharges used in this dataset are the following: 10708210[03/04/05/08/09/12/13/20/21/22/23/25/29].

\subsection{Neutral emissivity from Ly$_{\alpha}$ camera}
\label{subsec:lymid}
Radial profiles of $n_{0}$, $S_\mathrm{ion}$, $\Gamma_{D}$, and $D_\mathrm{eff}$ are reconstructed from a midplane Ly$_{\alpha}$ camera called LYMID, using the methodology outlined in Section \ref{subsec:neutral_inference}. The camera was installed in 2007 as the successor to an earlier Ly$_{\alpha}$ system \cite{boivin_effects_2000, boivin_high_2001}. It was located just below the midplane ($Z=-0.04$ m) and consisted of toroidal views. Its 20 chords were tangent to flux surfaces covering a region in the radial direction of just over 5 cm at the plasma edge. The LYMID diagnostic collected Ly$_{\alpha}$ light from the 121.6 nm vacuum ultra-violet light transition from the edge population of neutral deuterium atoms. This LYMID diagnostic recorded line-integrated brightness measurements from each of its viewing chords as a function of their tangency radius. We use the same tomographic inversion algorithm developed for LLAMA, a similar system on DIII-D \cite{rosenthal_1d_2021, odstrcil_optimized_2016} to invert the brightness profile. This yields $\epsilon_{\mathrm{Ly}_{\alpha}}$ as a function simply of radius, and no longer of toroidal angle, under the assumption of axisymmetric emission. No radial shift is applied to $\epsilon_{\mathrm{Ly}_{\alpha}}$. Instead, an error associated with the magnetic reconstruction is included, as delineated at the end of Section \ref{subsec:neutral_inference}.

\subsection{Edge Thomson scattering for plasma profile analysis}
\label{subsec:ETS}
In addition to neutral emissivity measurements, we use the ETS diagnostic, which was routinely used to diagnose the edge plasma, and in particular, the pedestal, on C-Mod. ETS measured both $n_{e}$ and $T_{e}$ in a region spanning 3 cm of the plasma edge. Measurements were located in the upper chamber near the crown of LSN plasmas, and could  diagnose $n_{e}$ and $T_{e}$ profiles with order millimeter resolution when mapped to the midplane \cite{hughes_high-resolution_2001}. In order to estimate gradients as well as to facilitate interpolation onto the radial coordinate basis of the neutral emissivity profile, a fit to both $n_{e}$ and $T_{e}$ is applied. It was determined that a hyperbolic tangent as in \cite{groebner_progress_2001}, but supplemented with polynomial terms to allow for flexibility in adjusting the gradient inside the pedestal and in the SOL, was appropriate to fit the plasma data. Equation \ref{eq:mtanh} shows the functional form of the modified hyperbolic tangent (mtanh).

\begin{equation}
    y(z) = \frac{1}{2}[h + b + (h - b) \frac{P_{1}(z)e^{z} + P_{2}(z)e^{-z}}{e^{z}+e^{-z}}]
    \label{eq:mtanh}
\end{equation}

$P_{1}(z)\,=\,1\,+\,C_{1}z+\,C_{2}z^{2}\,+\,C_{3}z^{3}$ and $P_{2}(z)\,=\,1+\,S_{1}z\,+\,S_{2}z^{2}$ are the core and SOL polynomials respectively, $z = 2 \frac{x_{0}-x}{\Delta}$ is the pedestal coordinate as a function of $x$, an arbitrary midplane coordinate, and $h$ and $b$ are the height and bottom of the pedestal, respectively. $x_{0}$ is the pedestal center, $\Delta$ is the pedestal width, $C_{1}$, $C_{2}$, and $C_{3}$ are the linear, quadratic, and cubic coefficients of the core polynomial, and $S_{1}$ and $S_{2}$ are the linear and quadratic coefficients of the SOL polynomial. Setting $C_{2} = C_{3} = S_{1} = S_{2} = 0$ recovers the form of the fit function used in \cite{groebner_progress_2001}. Since the mtanh has a closed form, the $n_{e}$ and $T_{e}$ gradients ($\nabla n_{e}$ and $\nabla T_{e}$) can be computed analytically using the fit coefficients. 

For ETS, we use the two-point model \cite{stangeby_pc_plasma_2000} to align $n_{e}$ and $T_{e}$ relative to the separatrix, $\psi_{n} = 1$. This model assumes that the power carried by the electrons in the SOL leaves the core entirely at the outer midplane and is transported to the divertor target via conduction only. It uses the Spitzer-Härm conductivity, which is proportional to the parallel temperature gradient in the SOL. Hence, the equality,

\begin{equation}
    -\frac{2}{7} \kappa_{0,e} \frac{\partial{T_{e}^{7/2}}}{\partial{x_{\parallel}}} = \frac{\frac{1}{2} P_\mathrm{net}}{2\pi R \lambda_{q} \frac{B_{\theta}}{B}}
    \label{eq:power_balance}
\end{equation}

can be integrated to find $T_{e}^\mathrm{sep}$ at the outer midplane (OMP) \cite{stangeby_pc_plasma_2000}. In this equation, $\kappa_{0,e}$ is the Spitzer-Härm conductivity, $R$ is the plasma major radius, $\lambda_{q}$ is the width of the heat flux channel, $B_{\theta}$ is the poloidal magnetic field. In the absence of measurements of the parallel heat flux width, we use the Brunner scaling for $\lambda_{q}$ \cite{brunner_high-resolution_2018}. We further assume negligible temperature at the divertor (i.e. $T_{e,\mathrm{target}}^{7/2} \ll T_{e,\mathrm{OMP}}^{7/2}$), as some target data available for these discharges indicate that $T_{e,\mathrm{target}}$ does not typically exceed 25 eV for even the highest values of $P_\mathrm{net}$. We also estimate the connection length required for the integral as $L_{\parallel} = q_{95} \pi R$, where $q_{95}$ is the safety factor at the 95\% flux surface. Finally, the factor of $\frac{1}{2}$ in front of $P_\mathrm{net}$ represents the assumption that half of the power crossing the separatrix is carried by the electrons (and half by the ions), an assumption that will be discussed further in the modeling setup in Section \ref{sec:solps}. With $T_{e,\mathrm{OMP}}^\mathrm{sep}$ in hand, we find the radial shift required to align the fit to the $T_{e}$ profile at $\psi_{n} = 1$ and apply the same shift to $n_{e}$ before using these shifted kinetic profiles to calculate the rate coefficients used for the neutral inferences.

\subsection{Inference of neutral density, ionization source, and particle flux}
\label{subsec:neutral_inference}
With $n_{e}$, $T_{e}$, and $\epsilon_{\mathrm{Ly}_{\alpha}}$ all mapped to the midplane, we make use of a collisional-radiative model to infer $n_{0}$, $S_\mathrm{ion}$, $\Gamma_{D}$, and an effective transport coefficient, $D_\mathrm{eff}$. A similar procedure has been employed previously, on C-Mod with D$_\mathrm{\alpha}$ measurements \cite{hughes_advances_2006}, and on both C-Mod and DIII-D with Ly$_\mathrm{\alpha}$ \cite{labombard_critical_2008, labombard_cross-field_2000, labombard_particle_2001, rosenthal_inference_2023, rosenthal_pedestal_2024}. Equations \ref{eq:n0} - \ref{eq:deff} summarize the procedure for calculating these quantities.

\begin{equation}
    n_{0} = \frac{\epsilon_{Ly-\alpha}}{\mathcal{P}(n_{e},T_{e})n_{e}}
    \label{eq:n0}
\end{equation}

\begin{equation}
    S_\mathrm{ion} = S_{H}(n_{e},T_{e}) n_{0}n_{e} = \epsilon_{Ly-\alpha}\frac{\mathcal{S}(n_{e}, T_{e})}{\mathcal{P}(n_{e},T_{e})}
    \label{eq:Sion}
\end{equation}

\begin{equation}
    \Gamma_{D,\perp}(r) = \int_{0}^{r} S_\mathrm{ion}(r')dr'
    \label{eq:flux}
\end{equation}

\begin{equation}
    D_\mathrm{eff} = \frac{\Gamma_{D,\perp}}{|\nabla n_{e}|}
    \label{eq:deff}
\end{equation}

$\mathcal{P}$ is the photon emissivity coefficient, $S_{H} = \frac{\mathcal{S}}{\mathcal{P}}$ is the ratio between $\mathcal{S}$, the effective ionization coefficient, and $\mathcal{P}$, the photon emissivity coefficient. Neutral and ionized molecular contributions are neglibible for Ly$_{\alpha}$ emission. These expressions highlight the implicit dependence of $S_\mathrm{ion}$ (and hence $\Gamma_{D,\perp}$) on $n_{e}$ and $T_{e}$.

\begin{figure*}
\centering
\includegraphics[width=2\columnwidth]{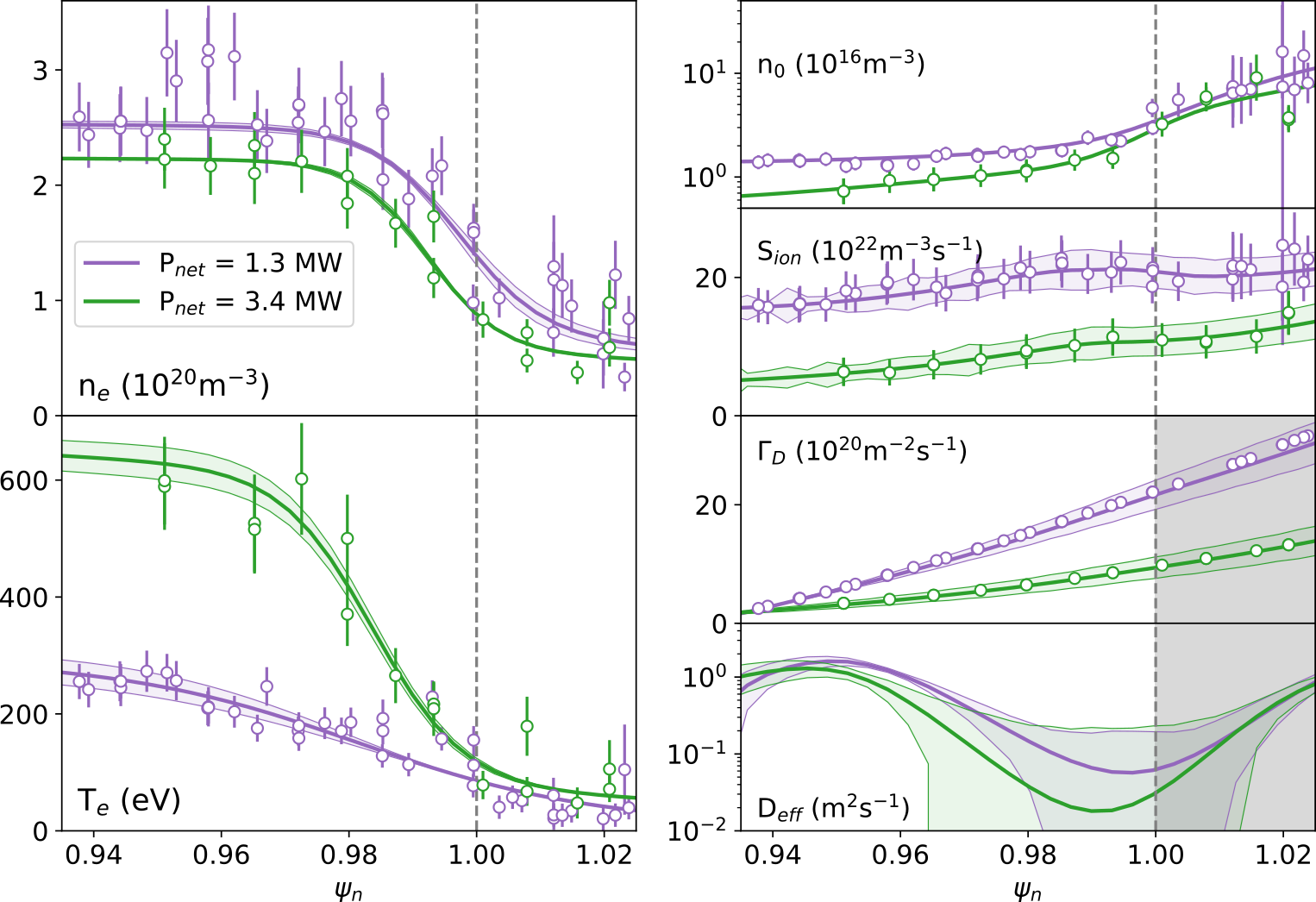}
\caption{Typical kinetic profiles as measured by ETS and a combination of LYMID and ETS measurements. The left panel shows profiles from ETS, with $n_{e}$ (top) and $T_{e}$ (bottom). Shown as open circles are the raw ETS points, including measurement uncertainties. Thick solid lines represent profile fits using the modified hyperbolic tangent fit as described in Section \ref{subsec:ETS}. Thinner solid lines and the shaded region they envelop represent uncertainties in the fit procedure, as outlined in the same section. The right panel shows those from combined ETS and LYMID analysis. From top to bottom are $n_{0}$, $S_\mathrm{ion}$, $\Gamma_{D}$, and $D_\mathrm{eff}$. The profiles in purple are at higher collisionality (higher $n_{e}$ and lower $T_{e}$), the relevance of which is explained in Section \ref{sec:os}. The ones in green are at lower collisionality (lower $n_{e}$ and higher $T_{e}$). The dotted vertical line in gray denotes the separatrix, located at $\psi_{n} = 1.0$. As mentioned in the text, $\Gamma_{D}$ and $D_\mathrm{eff}$ are only valid up to the separatrix. The profiles shown here correspond to cases simulated using SOLPS-ITER, described in Section \ref{sec:solps}.}
\label{fig:exp_profiles}
\end{figure*}

\begin{table}
\begin{center}
\caption{Ranges of values at the location of the top of the $n_{e}$ pedestal and at the separatrix for the discharges analyzed. $n_{e}$ and $T_{e}$ vary substantially at the separatrix, but $T_{e}$ primarily varies at the pedestal top. In contrast, all neutral quantities vary substantially across discharges and at both the pedestal and the separatrix.}
\label{tab:exp_params}
\begin{tabular}{l|c|c}
\hline
Parameter & $n_{e}$ pedestal top & Separatrix \\
\hline
\hline
$n_{e}$ (10$^{20}$ m$^{-3}$) & 1.9 -- 2.3 & 0.7 -- 1.9 \\
$T_{e}$ (eV) &  210 -- 670 & 80 - 150 \\
$n_{0}$ (10$^{16}$ m$^{-3}$) & 1.0 -- 2.1 & 1.9 -- 4.3 \\
$S_\mathrm{ion}$ (10$^{22}$ m$^{-3}$s$^{-1}$) & 7.8 -- 21 & 9.3 -- 23 \\
$\Gamma_{D}$ (10$^{20}$ m$^{-2}$s$^{-1}$) & 4.7 -- 18 & 8.2 -- 22 \\
$D_\mathrm{eff}$ (m$^{2}$s$^{-1}$) & 0.025 -- 0.13 & 0.016 -- 0.11 \\
\hline
\end{tabular}
\end{center}
\end{table}

Equations \ref{eq:n0} and \ref{eq:Sion} use a number of simplifying assumptions about the relevant atomic processes, which are outlined in \cite{rosenthal_inference_2023}. For Equation \ref{eq:flux}, we ensure a stationary plasma density time window and assume no poloidal asymmetries in $S_\mathrm{ion}$, allowing computation of the cross-field particle flux, $\Gamma_{D,\perp}$ by integrating Equation \ref{eq:Sion} in a slab geometry. For the remainder of the paper, we drop the $\perp$ subscript and express the cross-field particle flux as $\Gamma_{D}$. The assumption that there is negligible poloidal asymmetry in $S_\mathrm{ion}$ is likely the strongest simplifying assumption. Using two distinct Ly$_{\alpha}$ views, an in-out poloidal asymmetry has been observed in the ionization source on DIII-D \cite{rosenthal_inference_2023, laggner_absolute_2021}. LYMID does not distinctly view the inboard and outboard side, and an in-out source asymmetry may well be present on C-Mod as well. Regardless, past work has shown that C-Mod is dominated by main-chamber fueling \cite{umansky_comments_1998, labombard_cross-field_2000}. This lends confidence to using neutral inferences at the midplane to make conclusions about fueling and transport in the C-Mod pedestal. We test this assertion with EIRENE, a 3D neutral simulation code, described in Section \ref{sec:solps}, and observe good agreement with previous conclusions about main chamber fueling.

Figure \ref{fig:exp_profiles} shows typical profiles produced with this workflow. The panel on the left shows $n_{e}$ and $T_{e}$ measured by ETS as well as profile fits using Equation \ref{eq:mtanh}. The right panel of the figure shows $n_{0}$, $S_\mathrm{ion}$, $\Gamma_{D}$, and $D_\mathrm{eff}$, all computed from a combination of data from LYMID and ETS. Note that given the non-negligible contribution to ionization from parallel particle flux, $\Gamma_{D,\parallel}$, in the SOL, we only calculate $\Gamma_{D}$ and $D_\mathrm{eff}$ for $\psi_{n} \leq$ 1. Table \ref{tab:exp_params} summarizes the range in values for these profiles both at the pedestal top, defined as the point $\Delta$/2 radially inward of the center of the mtanh profile, where $\Delta$ is the width of the mtanh (as well as the pedestal). Only the radial location at the top of the $n_{e}$ pedestal is shown. Tabulated as well are values at the separatrix.

Uncertainties in the profiles in Figure \ref{fig:exp_profiles} have been computed by perturbing the experimental data points, refitting and recomputing associated quantities multiple times, and then computing statistics on the fits. The perturbations to the experimental points are sampled from a Gaussian distribution with a half-width equal to the experimental uncertainty of the ETS data point. This process is repeated $N$ times and the uncertainty is set to the standard deviation of the ensemble of $N$ perturbed fits. Through sensitivity analysis, it was determined that $N=100$ represented a good compromise between computational time and representation of the uncertainty. This technique enables uncertainty estimation not only in the absolute $n_{e}$ and $T_{e}$ profiles, but also in their gradients, as well as the neutral quantities computed from the plasma profiles. Furthermore, like proper weighted least squares including off-diagonal terms, this uncertainty estimation technique accounts for covariance between fit parameters. The technique is described in greater detail in \cite{rosenthal_pedestal_2024}.

As indicated earlier, we do not apply a shift to $\epsilon_\mathrm{Ly_{\alpha}}$, as the measurement is made at the midplane and there is no model to constrain the measurement relative to the separatrix. To account for this uncertainty, we compute an error associated with uncertainty in the position of the separatrix as indicated by the magnetic reconstruction computed with EFIT, $\Delta R_\mathrm{EFIT}$ = 2 mm. We thus estimate the error in $\epsilon_\mathrm{Ly_{\alpha}}$ from LYMID, $\Delta \epsilon_\mathrm{Ly_{\alpha}}$, by multiplying $\Delta R_\mathrm{EFIT}$ by the radial derivative of the emissivity profile, $\frac{\partial \epsilon_\mathrm{Ly_{\alpha}}}{\partial r}$, computed numerically, i.e. $\Delta \epsilon_\mathrm{Ly_{\alpha}} = \Delta R_\mathrm{EFIT}\frac{\partial \epsilon_\mathrm{Ly_{\alpha}}}{\partial r}$. This error is then added in quadrature to the quantities computed from Equations \ref{eq:n0} -- \ref{eq:deff}.

\section{Density pedestal stiffness}
\label{sec:fueling}

\subsection{The role of ionization in the pedestal}
\begin{figure}[h!]
\centering
\includegraphics[width=0.95\columnwidth]{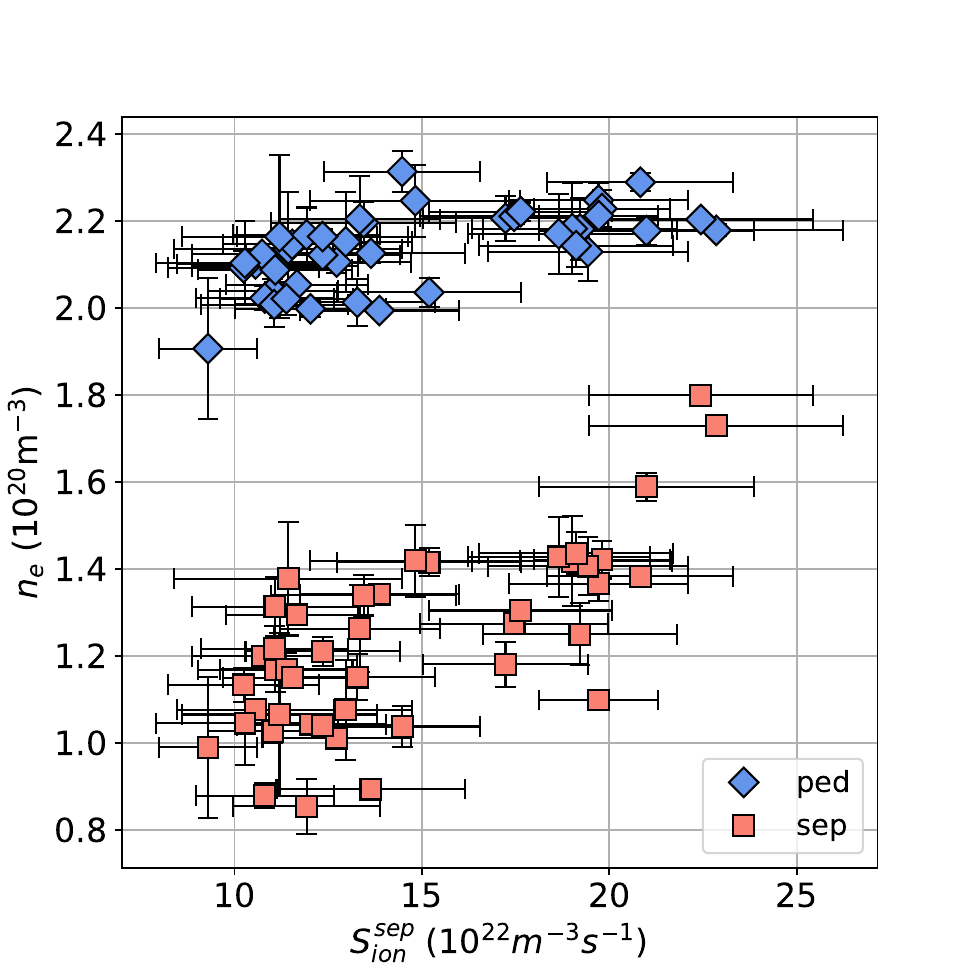}
\caption{Electron density plotted against volumetric neutral ionization rate at the outer midplane at the separatrix (light red squares) and at the top of the $n_{e}$ pedestal (light blue diamonds). While $n_{e}^\mathrm{sep}$ continues to increase as $S_\mathrm{ion}^\mathrm{sep}$ does, $n_{e}^\mathrm{ped}$ does not beyond a value of $n_{e}^\mathrm{ped} \approx 2.35 \times 10^{20}$ m$^{-3}$.}
\label{fig:n_vs_Sion}
\end{figure}

We use this methodology to interrogate the role of pedestal ionization in influencing the electron density pedestal height. Figure \ref{fig:n_vs_Sion} shows $n_{e}$ plotted against $S_\mathrm{ion}$ at the separatrix, $S_\mathrm{ion}^\mathrm{sep}$. At low values of separatrix ionization, $S_\mathrm{ion}^\mathrm{sep} \lesssim 15\times 10^{22}$ m$^{-3}$s$^{-1}$, $n_{e}$ varies weakly with ionization at the pedestal top and perhaps more strongly at the separatrix. As $S_\mathrm{ion}$ continues to grow past $S_\mathrm{ion}^\mathrm{sep} = 15\times 10^{22}$ m$^{-3}$s$^{-1}$, however, the trends diverge significantly. At the separatrix, $n_{e}$ continues to increase with $S_\mathrm{ion}$, while at the pedestal top there is clear insensitivity of the $n_{e}$ pedestal height to the ionization source. These EDA H-modes settle at a natural density of $n_{e}^\mathrm{ped} \approx 2.2 \times 10^{20}$ m$^{-3}$. Pedestal stiffness on Alcator C-Mod has been observed previously and may be a feature of high density EDAs \cite{hughes_edge_2007}. In that work, it was shown that varying the neutral source through external gas puffing had little effect on gradient scale lengths in the pedestal as well as the pedestal height. Though similar phenomenology is observed here, a key difference is that the H-modes currently analyzed received no external gas puffing after the L-H transition. Once the H-mode was achieved, pedestal fueling occurred solely through recycled neutrals. Pedestal stiffness in opaque plasmas may thus occur naturally, even in the absence of active gas fueling.

\subsection{Flux-gradient relationship in stiff profiles}
\begin{figure}[h!]
\centering
\includegraphics[width=0.9\columnwidth]{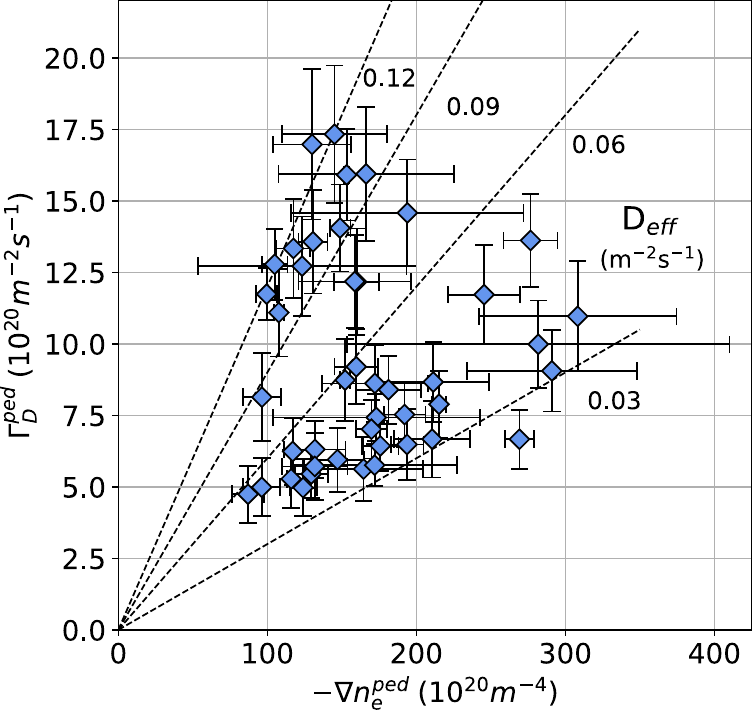}
\caption{Particle flux at the top of the pedestal plotted against density gradient at the same location. Even at large $\Gamma_{D}$, $-\nabla n_{e}$ does not increase much past $300\times10^{20}$ m$^{-4}$. The dashed lines show different slopes, corresponding to different values of $D_\mathrm{eff}$. The points lie at a large range of $D_\mathrm{eff}$, indicating inconsistency with a diffusive transport model using a constant $D$.}
\label{fig:flux_grad}
\end{figure}

It has often been assumed that $S_\mathrm{ion}$ determines $n_{e}$ \cite{groebner_role_2002, mahdavi_physics_2003}. With this framework, we might expect $\Gamma_{D}$, the integral of $S_\mathrm{ion}$, to be proportional to the gradient in the density. In other words, $\Gamma_{D}$ and $\nabla n_{e}$ would be linearly dependent, just as in a classical Fickian diffusive system. Figure \ref{fig:flux_grad}, however, shows that the experimentally inferred $\Gamma_{D}$ from the Ly$_{\alpha}$ camera, calculated at the top of the pedestal does not grow linearly with $\nabla n_{e}$ at the same location. For $\nabla n_{e} \approx -150 \times 10^{20}$ m$^{-4}$, $\Gamma_{D}$ can vary significantly, anywhere from $6 \times 10^{20}$ m$^{-2}$s$^{-1}$ to $16 \times 10^{20}$ m$^{-2}$s$^{-1}$. This is not surprising, given the saturation of $n_{e}^\mathrm{ped}$ for vastly different $S_\mathrm{ion}$ shown in Figure \ref{fig:n_vs_Sion}. 

Figure \ref{fig:flux_grad} indicates that there may be a particular gradient beyond which further ionization becomes ineffective in modifying the density profile, yielding a stiff pedestal. The peak $\nabla n_{e}$ occurs in the mid-range of $\Gamma_{D}$ values, at $\sim$10 $\times 10^{20}$ m$^{-2}$s$^{-1}$. This is consistent with the rapid growth of $n_{e}^\mathrm{sep}$ at fixed $n_{e}^\mathrm{ped}$ shown in Figure \ref{fig:n_vs_Sion}. Discharges at low $\Gamma_{D}^\mathrm{ped}$ correspond to those at low $S_\mathrm{ion}^\mathrm{sep}$ and low $n_{e}^\mathrm{sep}$. Those at mid $\Gamma_{D}^\mathrm{ped}$ and largest $\nabla n_{e}^\mathrm{ped}$ correspond to discharges at $S_\mathrm{ion}^\mathrm{sep} \approx 15 \times$ 10$^{22}$ m$^{-3}$s$^{-1}$ and intermediate $n_{e}^\mathrm{sep}$. Finally, the discharges at high $\Gamma_{D}^\mathrm{ped}$ but reduced $\nabla n_{e}^\mathrm{ped}$ (relative to those at mid $\Gamma_{D}^\mathrm{ped}$) correspond to those at highest $S_\mathrm{ion}^\mathrm{sep}$. These are the discharges with increasing $n_{e}^\mathrm{sep}$ but nearly fixed $n_{e}^\mathrm{ped}$. Generally also, discharges at low $S_\mathrm{ion}$ and low $\Gamma_{D}$ correspond to those at high $P_\mathrm{net}$ and vice versa, as suggested by Figure \ref{fig:pnet}.

It is evident that these pedestals cannot be fully described with a purely diffusive transport model of constant $D$. Since $S_\mathrm{ion}$, and thus $\Gamma_{D}$, varies significantly at relatively constant $n_{e}^\mathrm{ped}$, it is unlikely that just adding a convective term, proportional to $n_{e}^\mathrm{ped}$, to the transport model might explain the large changes to $\Gamma_{D}$ in these pedestals. Instead, it could be that a threshold based model, like that describing heat transport near critical gradients in the core \cite{hillesheim_observation_2013}, may be necessary for understanding pedestal particle transport. Such a phenomenon in the particle channel has been previously suggested on Alcator C-Mod in the near-SOL for ohmic L-modes \cite{labombard_critical_2008} and in the pedestal of EDA H-modes undergoing strong puffing \cite{hughes_edge_2007}.  

In the absence of such a model, we opt for using the quotient of the inferred particle flux and the density gradient to obtain an effective particle diffusivity, $D_\mathrm{eff}$, calculated using Equation \ref{eq:deff}. While not fully descriptive of the physics of the edge, $D_\mathrm{eff}$ provides a point of comparison for analyzing differences in transport between discharges and as a function of different edge conditions. Experimental inferences of $D_\mathrm{eff}$ have been previously made on C-Mod using $\Gamma_{D}$ inferred from the workflow in Section \ref{subsec:neutral_inference} \cite{hughes_advances_2006, labombard_cross-field_2000}, although never within the pedestal with both LYMID and ETS simultaneously. An important caveat of this steady-state analysis is that in using a single coefficient, one cannot distinguish different types of particle transport, as becomes possible using time-dependent analysis. On DIII-D, for example, recent experiments have sought to calculate both a diffusive coefficient, $D$, and a convective velocity coefficient, $v$, simultaneously using LLAMA \cite{rosenthal_inference_2023}. To do this requires perturbing the edge slightly so as to recover $\frac{\partial n}{\partial t}$ without substantially changing the intrinsic transport properties. 

For these stationary discharges, we opt for $D_\mathrm{eff}$, and add lines of constant $D_\mathrm{eff}$ to Figure \ref{fig:flux_grad} to demonstrate the large variation in this proxy for transport in these plasmas. As shown in the rightmost panel of Figure \ref{fig:pnet} and now apparent from Figures \ref{fig:n_vs_Sion} and \ref{fig:flux_grad}, the rapid growth in $S_\mathrm{ion}$ and $\Gamma_{D}$ at low $P_\mathrm{net}$, as well as the saturation of $n_{e}^\mathrm{ped}$, occurs along with a rapid growth of $D_\mathrm{eff}^\mathrm{ped}$ below the critical value of $P_\mathrm{net} \approx 2.3$ MW.

\section{Pedestal operational space and transport drive}
\label{sec:os}

\begin{figure}
\includegraphics[width=\columnwidth]{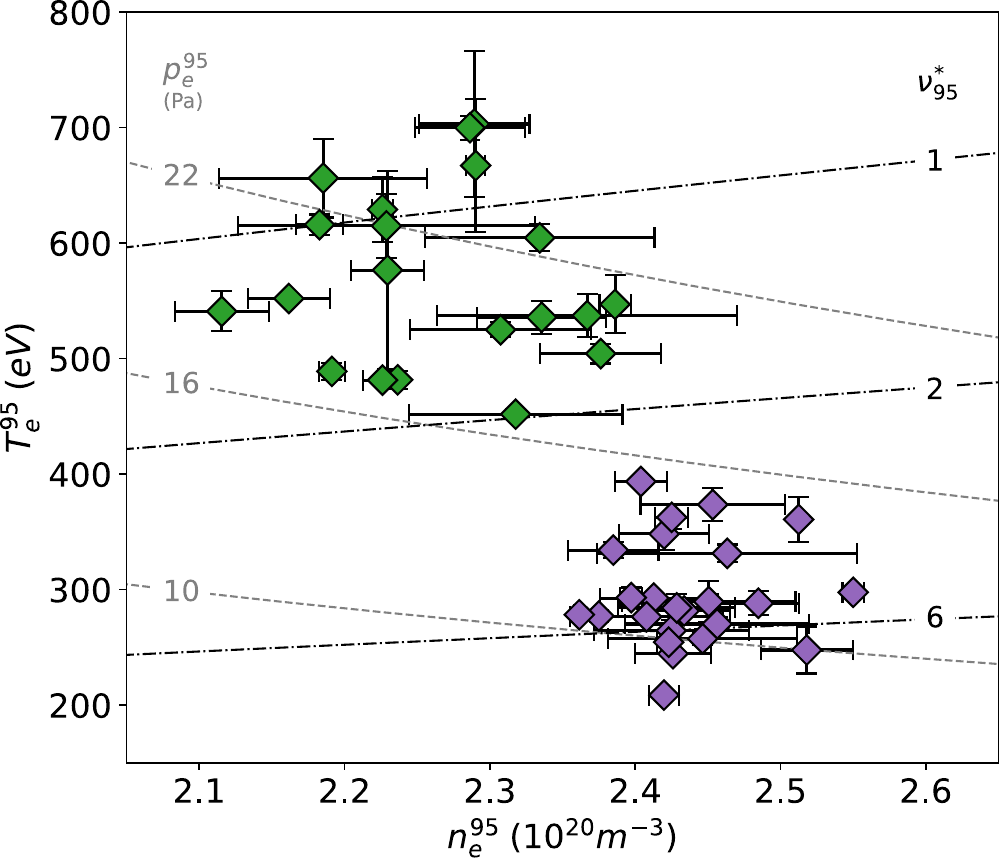}
\caption{Pedestal top operational space in terms of ($n_{e}$,$T_{e}$) at $\psi_{n}$ = 0.95. Purple diamonds are at high $n_{e}$ and low $T_{e}$, while green diamonds are at low $n_{e}$ and high $T_{e}$. Isobars (gray, dashed) and constant collisionality contours (black, dash-dotted) are included.}
\label{fig:ped_os}
\end{figure}

Particle transport and fueling depend on and affect $T_{e}$ as well as $n_{e}$. Neutral transport resulting from atomic processes has a strong dependence on $T_{e}$, and intrinsic changes to plasma transport modify and are modified by kinetic profiles, including $T_{e}$ \cite{hughes_advances_2006, mordijck_overview_2020}. Figure \ref{fig:exp_profiles} shows characteristic $n_{e}$ and $T_{e}$ pedestal profiles at low and high $P_\mathrm{net}$, shown in green and purple respectively. There is a clear drop in $T_{e}^\mathrm{ped}$ at low $P_\mathrm{net}$. Figure \ref{fig:ped_os} shows the phase space of these plasmas in terms of $n_{e}$ and $T_{e}$, at $\psi_{n} = 0.95$, ($n_{e}^{95}$, $T_{e}^{95}$). This location is slightly inside the pedestal top where gradients are considerably smaller than in the respective pedestals themselves, so that generally, they are not very different from the value at the top of the pedestal, and so, ($n_{e}^\mathrm{ped}, T_{e}^\mathrm{ped}) \sim (n_{e}^{95}, T_{e}^{95}$). We use this location instead of the top of the pedestal defined by the mtanh to facilitate the use of collisionality as explained below in Section \ref{subsec:p_nu}. Indeed $T_{e}^{95}$ varies dramatically in this set of experiments. In fact, within error bars, these pedestals cluster at opposite corners of the operational space - at high $n_{e}^{95}$ and low $T_{e}^{95}$, and low $n_{e}^{95}$ and high $T_{e}^{95}$. For the remainder of the paper, we distinguish the former using purple and the latter using green. Purple markers are also those at low $P_\mathrm{net}$ and high $\Gamma_{D}$, while green ones at higher $P_\mathrm{net}$ also maintain lower $\Gamma_{D}$. We make a more quantitative distinction between these in the next section.

\begin{figure*}
\centering
\includegraphics[width=1.7\columnwidth]{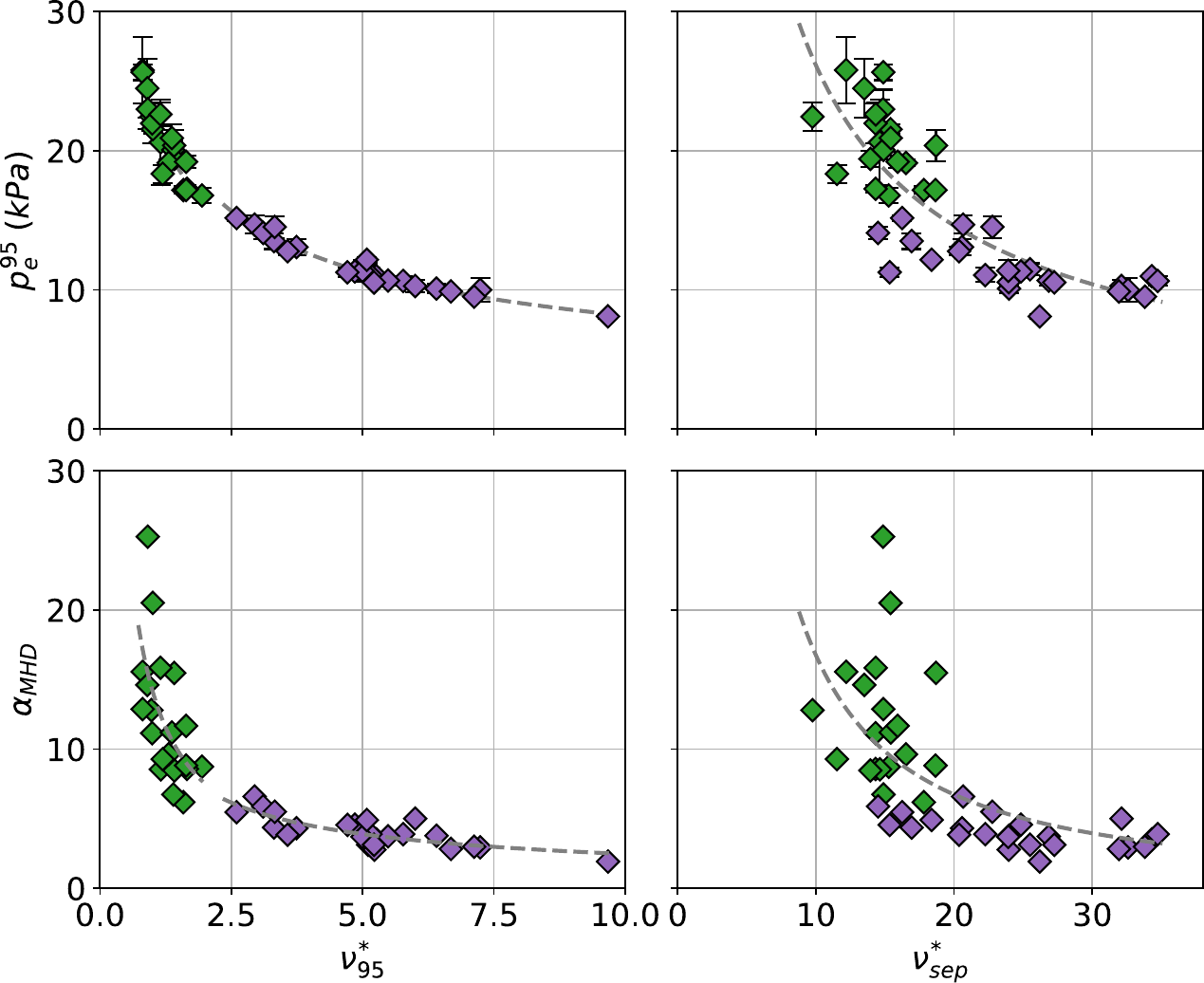}
\caption{Pressure at $\psi_{n} = 0.95$ (top) and maximum normalized pressure gradient (bottom) plotted against collisionality at $\psi_{n} = 0.95$ (left) and at the separatrix (right). In both plots at left, there is a clear dependence of $p_{e}^{95}$ and $\alpha_\mathrm{MHD}$ on $\nu^{*}_{95}$, whereby higher $\nu^{*}_{95}$ inhibits large pedestal pressure or pressure gradient. Plotting against $\nu^{*}_\mathrm{sep}$ shows similar dependence, but with a less abrupt transition in $\nu^{*}$. Shown also are power law fits, to both high and low $\nu^{*}_{95}$ points separately for the plots on the left, and across the full range of $\nu^{*}_\mathrm{sep}$ for those on the right. For the plot on the top left, both high and low $\nu^{*}$ discharges trend in the same way, with $p_{e}^{95} \sim (\nu^{*}_{95})^{-0.5}$. For the plot on the bottom, $\alpha_\mathrm{MHD}$ decreases more strongly at low $\nu^{*}$ and more weakly at high $\nu^{*}$. Their trends for low and high $\nu^{*}$ are $\alpha_\mathrm{MHD} \sim (\nu^{*}_{95})^{-0.9}$ and $\alpha_\mathrm{MHD} \sim (\nu^{*}_{95})^{-0.7}$, respectively. When regressing against $\nu^{*}_\mathrm{sep}$, as in the plots on the right, a stronger dependence is found for both $p_{e}^{95}$ and $\alpha_\mathrm{MHD}$, with $p_{e}^{95} \sim (\nu^{*}_\mathrm{sep})^{-0.8}$ and $\alpha_\mathrm{MHD} \sim (\nu^{*}_\mathrm{sep})^{-1.3}$.}
\label{fig:pressure_pedestal}
\end{figure*}

\subsection{Pressure and collisionality}
\label{subsec:p_nu}
Figure \ref{fig:ped_os} includes both isobars in gray, dashed curves and constant collisionality contours (CCC) in and black, dash-dotted curves. The curves are both calculated at $\psi_{n} = 0.95$. The isobars are calculated using $p_{e} = n_{e}T_{e}$ and the CCCs are calculated using $\nu^{*}_{95} = \frac{q_{95}R_{0}\nu_{ei}}{\epsilon^{3/2}v_{\mathrm{th},e}}$, where $q_{95}$ is the safety factor at $\psi_{n} = 0.95$, $R_{0}$ is the plasma major radius, $\nu_{ei}$ is the electron-ion collision frequency, $\epsilon$ is the inverse aspect ratio, and $v_{\mathrm{th},e}$ is the electron thermal velocity. $\nu_{e,i}$ depends on $Z_\mathrm{eff}$, the effective charge of the plasma, which is a difficult quantity to measure on C-Mod. Since these plasmas did not have extrinsic impurity seeding, $Z_\mathrm{eff}$ is taken to be 1.4, a common assumption for relatively pure plasmas on C-Mod. Since all of these plasmas are at similar $B_{t}$ and $I_{P}$, there is little variation in $q_{95}$, so we calculate the CCCs from its average, $\langle q_{95} \rangle = 4.7$. From Figure \ref{fig:ped_os}, it is apparent that these pedestals range substantially both in $p_{e}^{95}$ and $\nu^{*}_{95}$. Pedestals with low $D_\mathrm{eff}$ can reach up pressures up to $p_{e}^{95} = 26.5$\,kPa, more than twice that of pedestals with high $D_\mathrm{eff}$. Conversely, transport-ridden pedestals at low $p_{e}^{95}$ can reach collisionalities close to $\nu^{*}_{95} = 10$. Despite rather continuous changes in heat and particle sources as seen in Figure \ref{fig:pnet}, there appears to be a quite clear separation in $\nu^{*}_{95}$. Green points are those at $\nu^{*}_{95} < 2$ and the purple points are those at $\nu^{*}_{95} > 2$.

Plotting $p_{e}^{95}$ against $\nu^{*}_{95}$ as in the top left panel of Figure \ref{fig:pressure_pedestal} yields a relatively scatter-free correlation. Clearly, collisionality imposes rather strict regulation on the pedestal pressure, with $p_{e}^{95} \sim (\nu^{*}_{95})^{-0.5}$ across both groups of data. In fact, it seems that collisionality not only affects the attainable $p_{e}^\mathrm{ped}$, but \emph{also} the pressure pedestal gradient, $\nabla p^\mathrm{ped}$. Previous pedestal analysis on C-Mod has observed that the dimensionless pressure gradient is linked to the pedestal collisionality \cite{hughes_edge_2007}. The right panel of Figure \ref{fig:pressure_pedestal} supports this, showing the normalized pressure gradient, $\alpha_\mathrm{MHD}$, plotted against $\nu^{*}_{95}$, where $\alpha_\mathrm{MHD} = \frac{2\mu_{0}q_{95}^{2}R_{0}}{B_{T}^{2}}\nabla p$. We use the maximum $\nabla p$ in the edge, not strictly co-located with the center of the $n_{e}$ pedestal. Since $T_{i}$ measurements were not available for these discharges, we take $p_{i} = p_{e}$ and hence, set $\nabla p = 2\nabla p_{e}$. Furthermore, the expression used here is for a circular cross section and does not account for geometric corrections. Since the plasma shape is constant in all discharges shown here and the values of $\alpha_\mathrm{MHD}$ reported are not meant to make direct contact with a stability analysis but rather motivate the governing physics, we opt for this simplified version. Note that most parameters in $\alpha_\mathrm{MHD}$ are relatively constant across these discharges, with the exception of $\nabla p_{e}$. In other words, changes in $\alpha_\mathrm{MHD}$ are primarily changes in $\nabla p_{e}$ in this dataset. 

As with $p_{e}$, $\nabla p_{e}$ is very tightly regulated by collisionality, decreasing even more strongly with $\nu^{*}_{95}$. At low $\nu^{*}_{95}$, $\alpha_\mathrm{MHD} \sim (\nu^{*}_{95})^{-0.9}$. At high $\nu^{*}_{95}$, the dependence is slightly weaker, with $\alpha_\mathrm{MHD} \sim (\nu^{*}_{95})^{-0.7}$. It may be that as $\nu^{*}_{95}$ increases further, $\alpha_\mathrm{MHD}$ reaches some minimum value required to sustain an H-mode. When comparing against collisionality at the separatrix, $\nu^{*}_\mathrm{sep}$, both $p_{e}^{95}$ and $\alpha_\mathrm{MHD}$ trend more strongly across the range in $\nu^{*}_\mathrm{sep}$, with $p_{e}^{95} \sim (\nu^{*}_\mathrm{sep})^{-0.8}$ and $\alpha_\mathrm{MHD} \sim (\nu^{*}_\mathrm{sep})^{-1.3}$. They exhibit greater scatter, however, perhaps as a result of the radial separation in the parameters on the ordinate and abscissa.

Weakening of $\nabla p_{e}$ can be a symptom of differences in the position of the $n_{e}$ pedestal relative to the $T_{e}$ pedestal. Since $\nabla p_{e}$ consists of both $\nabla n_{e}$ and $\nabla T_{e}$, its value is strongly related to the position of each gradient. Misalignment in the positions of maximum gradient strongly impact $\alpha_\mathrm{MHD}$. The relative shift of the two profiles has been previously observed and is thought to be a fueling effect, where the peak of the ionization moves radially outward, increasing density and locally cooling the plasma \cite{hughes_edge_2007, beurskens_h-mode_2011, dunne_role_2017, wang_effects_2018, stefanikova_effect_2018, frassinetti_role_2019}. Of course, $\nabla p_{e}$ can also decrease if either of its constituent gradients decreases. In this dataset, discharges at low $\nu^{*}$ (green) have a relative density and temperature shift, $\Delta R_\mathrm{n-T}^\mathrm{ped} = \Delta R_\mathrm{n^\mathrm{ped}} - R_\mathrm{T^\mathrm{ped}}$ = 1.4 mm, whereas those at high $\nu^{*}$ (purple) have a relative shift $\Delta R_\mathrm{n-T}^\mathrm{ped}$ = 2.0 mm. On the other hand, low $\nu^{*}$ discharges have on average, a maximum $\nabla T_{e} \approx$ 123 keVm$^{-1}$, whereas high $\nu^{*}$ discharges only have a maximum $\nabla T_{e} \approx$ 32 keVm$^{-1}$ on average. When multiplied by $\frac{R}{T_{e}^\mathrm{ped}}$ as in \cite{luda_validation_2023}, giving an estimate of the $T_{e}$ gradient scale length, $L_{T_{e}}$ normalized by machine major radius, $R$, the discharges at low $\nu^{*}$ have an average $\frac{R}{L_{T_{e}}} \approx$ 162, while those at high $\nu^{*}$ have an average $\frac{R}{L_{T_{e}}} \approx$ 76, less than half of those at low $\nu^{*}$. While the gradients are computed at a different radial location than in \cite{luda_validation_2023}, it is informative to note that multi-machine scalings may diverge in these very high $\nu^{*}$ pedestals. Indeed, it appears that as $\nu^{*}_{95}$ increases, it is primarily the $\nabla T_{e}$ weakening that is responsible for the weakening of $\alpha_\mathrm{MHD}$ and a threat to fusion performance, reliant on a strong $\nabla p$.

\subsection{Transport variation with pedestal parameters}
\label{subsec:transport_pedestal}
\begin{figure*}
\centering
\includegraphics[width=1.6\columnwidth]{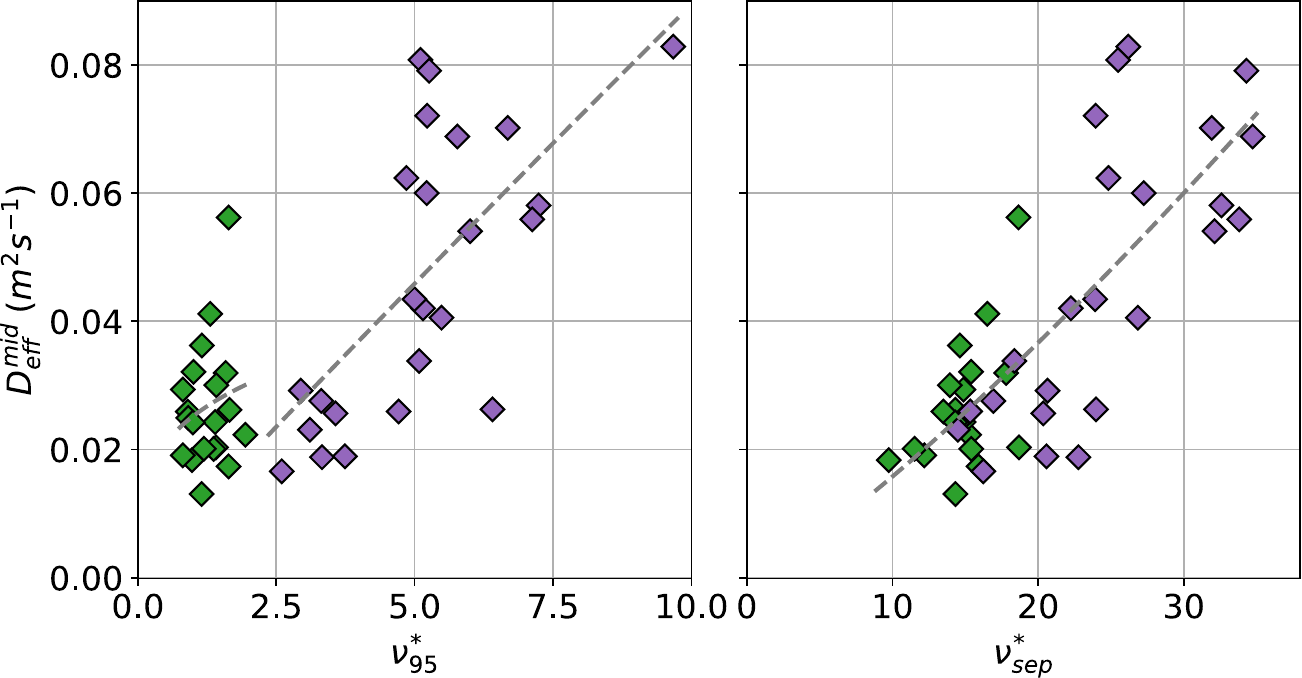}
\caption{Inferred $D_\mathrm{eff}$ at mid-pedestal plotted against collisionality at $\psi_{n} = 0.95$ (left) and the separatrix (right), both using $q_{95}$. Dashed lines represent power law fits to green and purple discharges independently. There is a clear separation in collisionality at the top of the pedestal. For low values of $\nu^{*}$, there is an unconvincing trend of $D_\mathrm{eff}^\mathrm{mid} \sim (\nu^{*}_{95})^{0.3}$. At high $\nu^{*}_{95}$, the trend is clearer - $D_\mathrm{eff}^\mathrm{mid} \sim (\nu^{*}_{95})^{1.0}$. When plotting against $\nu^{*}$ at the separatrix, the separation on the abscissa disappears and both sets of points see even stronger trends in the mid-pedestal particle transport of $D_\mathrm{eff}^\mathrm{mid} \sim (\nu^{*}_\mathrm{sep})^{1.2}$.}
\label{fig:deff_nu}
\end{figure*}

Simultaneous changes in $p_{e}$, $\nabla p_{e}$ and $\nu^{*}$ make it challenging to discern causality. Regardless, it is clear from Section \ref{sec:fueling} that there are large changes to $D_\mathrm{eff}$, which appear to affect pedestal character substantially, most notably by limiting $n_{e}$. Observing that the discharges colored in purple are also more collisional, it is natural to question whether pedestal collisionality may be responsible for driving this increased transport. To test this hypothesis, the left panel of Figure \ref{fig:deff_nu} shows how $D_\mathrm{eff}$ at mid-pedestal, $D_\mathrm{eff}^\mathrm{mid}$, varies with $\nu^{*}_{95}$. Not surprisingly, there is an increase in $D_\mathrm{eff}^\mathrm{mid}$ as $\nu^{*}_{95}$ increases. Regardless, the increase in $D_\mathrm{eff}$ with $\nu^{*}_{95}$ is evident only for $\nu^{*}_{95} > 2$. The right panel of Figure \ref{fig:deff_nu} shows the same metric for transport on the ordinate, but now as a function of $\nu^{*}$ calculated at the separatrix. At the separatrix, $\nu^{*}$ can be almost an order of magnitude higher than at the pedestal top. Given the large gradients in the H-mode pedestal this variation is not surprising. Plotting transport against $\nu^{*}_\mathrm{sep}$ yields a smoother correlation between transport and collisionality, implying that the separatrix conditions may more strongly influence pedestal gradients in these EDA H-modes than those at the pedestal top.

\subsection{Influence of separatrix condition on pedestal gradients}
That the separatrix encodes information about gradients and confinement inside the separatrix is not a novel concept. It has been previously observed on ohmic discharges on C-Mod that conditions around the separatrix strongly influence pressure gradients and the subsequent accumulation of plasma pressure inside of the separatrix \cite{labombard_evidence_2005}. Using a similar framework of electromagnetic fluid drift turbulence, it has been shown on ASDEX Upgrade (AUG) that the ($n_{e}$, $T_{e}$) operational space at the separatrix may strongly influence the overall plasma confinement regime as well as proximity to density limit-induced plasma disruption \cite{Eich_2021, manz_power_2023}. These works apply interchange-drift-Alfvén turbulence theory, identifying $\alpha_{t}$ as an important turbulence control parameter. The full expression for $\alpha_{t}$ can be found in Equation 10 of \cite{Eich_2020}, but for these C-Mod plasmas with $Z_\mathrm{eff}$ = 1.4, it can be approximated by Equation \ref{eq:alpha_t}.

\begin{equation}
    \alpha_{t} = 2.98 \times 10^{-18} R_\mathrm{geo}\hat{q}_\mathrm{cyl}^{2} \frac{n_{e}}{T_{e}^{2}} Z_\mathrm{eff}
    \label{eq:alpha_t}
\end{equation}

Here, $R_\mathrm{geo}$ is the geometric major radius, taken to be equal to the device major radius, $R_{0}$, and $\hat{q}_\mathrm{cyl}$ is the cylindrical safety factor, calculated according to $\hat{q}_\mathrm{cyl} = \frac{B_{t}}{B_{p}} \times \frac{\hat{\kappa}}{R_\mathrm{geo}/a_\mathrm{geo}}$, where $a_\mathrm{geo}$ is the geometric minor radius, taken equal to the device minor radius, and $\hat{\kappa}$ is the effective elongation, defined in \cite{Eich_2021} in terms of triangularity and geometric elongation, which in this work is taken as the elongation calculated by EFIT.

\begin{figure}[h!]
\centering
\includegraphics[width=0.85\columnwidth]{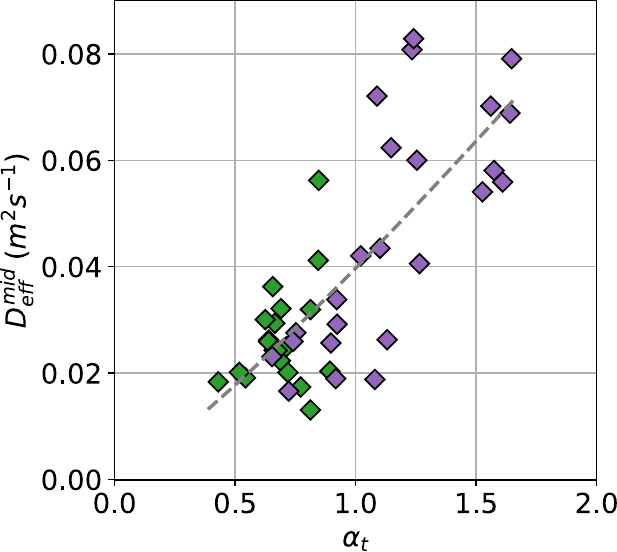}
\caption{Inferred $D_\mathrm{eff}$ at mid-pedestal plotted now against $\alpha_{t}$, a collisionality-like turbulence control parameter in interchange-DALF theory. It is suggested that when $\alpha_{t}$ reaches unity, edge turbulence transitions from DALF to interchange. The trend across both regimes is the same as with $\nu^{*}_\mathrm{sep}$, with $D_\mathrm{eff}^\mathrm{mid} \sim \alpha_{t}^{1.2}$.}
\label{fig:deff_alphat}
\end{figure}

It is observed that large values of $\alpha_{t}$ can widen the pressure gradient scale length ($\lambda_{p}$) at the separatrix \cite{Eich_2020}. Note that this parameter has the same form as the previously plotted $\nu^{*}$, but includes a quadratic rather than linear dependence on the safety factor, $q$. The importance of this additional factor on $q$ was noticed earlier as well in the work on C-Mod \cite{labombard_evidence_2005}. There, it was shown that $\lambda_{p}$ organized better across different values of $q_{95}$ when including this additional factor of $q$. In the current work, the plasmas are at similar $q_{95}$, so little re-organization of data occurs when plotting against $\alpha_{t}$ instead of $\nu^{*}$. Doing so, however, points to a possible physical explanation. Figure \ref{fig:deff_alphat} indicates that the transition between the high pressure and low pressure pedestals corresponds to values approaching and exceeding unity, the range in which resistive ballooning mode (RBM) turbulence is thought to exert strong influence over drift-wave (DW) in \cite{Eich_2021}. As the separatrix gets denser and colder and $\alpha_{t}$ approaches unity, RBMs may begin to drive particle transport, substantially increasing $D_\mathrm{eff}$. 

\section{Neutral modeling and interpretive transport solutions with SOLPS-ITER}
\label{sec:solps}

The technique outlined in Section \ref{sec:experiment} is powerful in that it can be employed across an entire database of shots with high quality Ly$_{\alpha}$ and ETS data. These are 1D measurements, however, so any conclusions about transport hinge on assumptions about the poloidal distribution of neutrals. If $\Gamma_{D}$ estimated at the outer midplane (OMP), $\Gamma_{D}^\mathrm{OMP}$, is significantly different its flux-surface averaged value, $\Gamma_{D}^\mathrm{FSA}$, the experimentally inferred $D_\mathrm{eff}$ may not be wholly indicative of transport in the pedestal across its poloidal extent. Experimental estimates of $D_\mathrm{eff}$ made here assume that variation in the poloidal fueling profiles between shots is small and that generally, this peaks at or near the poloidal location observed by the Ly$_{\alpha}$ camera (here, the OMP). Furthermore, without some interpretive power balance analysis using a transport code, we can only investigate the particle transport and know little about changes to the electron or ion thermal transport. To gain insight into thermal transport, compare with our inferences of particle transport from 1D analysis, and learn about poloidal distribution of neutrals, we choose certain discharges from these experiments to simulate with SOLPS-ITER, a code suite used extensively for edge modeling of plasma and neutrals. SOLPS-ITER couples a 2D multi-fluid plasma transport code, B2.5, with a 3D kinetic Monte Carlo neutral transport solver, EIRENE \cite{wiesen_new_2015}. In the current study, we focus primarily on the OMP and constrain the plasma and neutral models there. The simulations are performed for only deuterium, D, ion species.  

\subsection{Plasma constraints}
\label{sec:plasma}

\subsubsection{Boundary conditions}
\label{subsubsec:bcs}
For the plasma solution, we reproduce $n_{e}$ and $T_{e}$ from the top of the pedestal to the near-SOL using profile fits to the ETS measurements. At the core boundary, we prescribe input power and particle fluxes. We take $P_\mathrm{net}$ from experiment and assume energy equipartition, providing $\frac{1}{2}P_\mathrm{net}$ to each the electron and the ion populations. While the exact partition of energy in the edge can vary in different plasmas, the assumption of equal heat fluxes falls within a range of values estimated from TRANSP runs of a number of C-Mod plasmas \cite{schmidtmayr_univ-doz_nodate}. Applying $P_\mathrm{net}$ at the innermost flux surface of the B2.5 plasma grid rather than at the separatrix assumes negligible radiated power in the plasma annulus about 1 cm thick between this flux surface and the separatrix. For the particle flux, we set $\Gamma = 0$. Alcator C-Mod had no core particle source, neither via neutral beam injection (NBI), nor pellet-fueling. Any non-zero $\Gamma$ at the innermost boundary of the B2.5 grid comes entirely from neutrals recycled at its walls, assumed to be negligible. This is checked a posteriori to be a valid assumption, as shown below in Section \ref{subsec:neutrals_solps}. 

Because of the small $\Delta R_\mathrm{sep} \approx 5$ mm, a disconnected DN grid is used, yielding four targets in the simulation domain. At each of these, sheath BCs are used, set via the usual modified Bohm criterion. For the SOL and private flux region (PFR), BCs are set to the ``leakage" type, instead of the more conventional ``decay length" type. The choice for these BCs is important for reproducing the details of the neutral profiles and are outlined in Section \ref{subsec:neutrals_solps}. Finally, motivated by a study for the importance of using heat flux limiters to match simulations to experimental C-Mod divertor temperature measurements \cite{brunner_assessment_2013}, flux limiters for the thermal parallel transport are included in the simulations, using a value of 0.15 for both ions and electrons.

\subsubsection{Transport coefficients}
\label{subsec:transport_coefs}
For particle transport, we assume purely diffusive transport, specifying the particle density diffusion coefficient, $D_{n}$. For energy, we use thermal diffusivity coefficients, $\chi_{e}$ for the electrons, and $\chi_{i}$ for the ions. To reproduce the steep gradients in these H-mode edges, it is necessary to apply a transport well by using a radially-varying profile in all three channels. We ignore poloidal variation of transport coefficients in the core and specify transport profiles to reproduce $n_{e}$ and $T_{e}$ measurements from ETS, at the OMP. This is likely a source of error, as previous work on C-Mod has shown that ballooning-like transport drives larger plasma fluxes to the outboard rather than inboard side of the core \cite{labombard_particle_2001}. As a result, the $D_\mathrm{eff}$ reported in this work may represent an underestimation of the transport on the LFS and an overestimation on the HFS. We further assume negligible poloidal variation in radial transport along open field lines in the SOL, such that the same transport coefficient profile at the midplane SOL is used at the divertor targets. Finally, we use flat transport coefficients in the PFR: $D_{n}$ = 5 $\times$ 10$^{-2}$ m$^{-2}$s$^{-1}$ and $\chi_{e}=\chi_{i}$ = 7 $\times$ 10$^{-1}$ m$^{-2}$s$^{-1}$.

\begin{figure}
\centering
\includegraphics[width=\columnwidth]{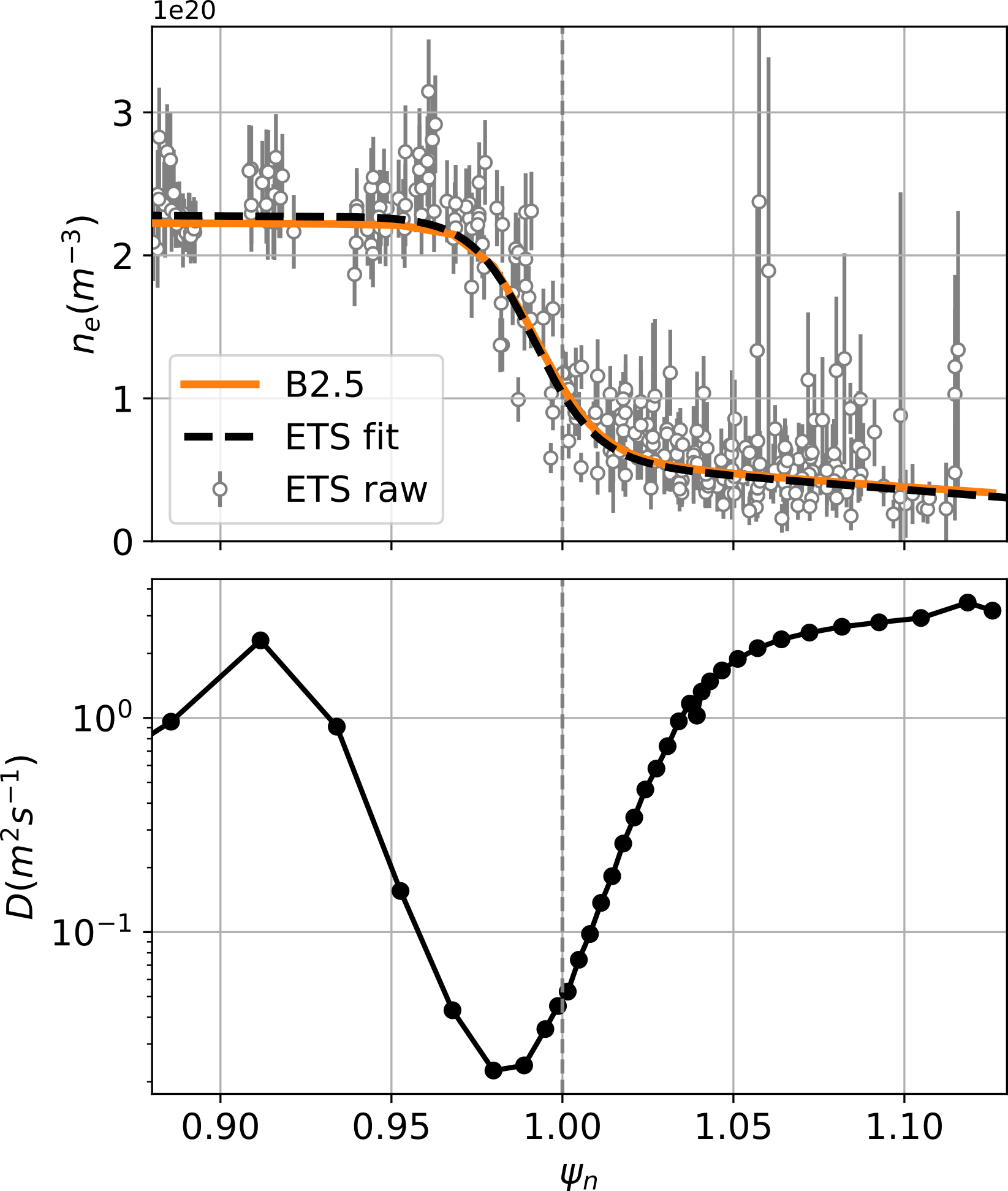}
\caption{Electron density (top) and particle diffusion coefficient (bottom) from a converged SOLPS simulation. The top plot shows the experimental ETS data points and their errorbars in open, gray circles, as well as the best mtanh fit (dashed black line) computed from the least-squares minimization procedure outlined in Section \ref{subsec:ETS}. The orange line shows the resulting $n_{e}$ after a set of iterations to determine the $D(\psi_{n})$ profile, plotted below, that would best reproduce the experimental $n_{e}(\psi_{n})$, while simultaneously matching $n_{0}(\psi_{n})$.}
\label{fig:ne_dn_solps_profs}
\end{figure}

\begin{figure}[h!]
\centering
\includegraphics[width=\columnwidth]{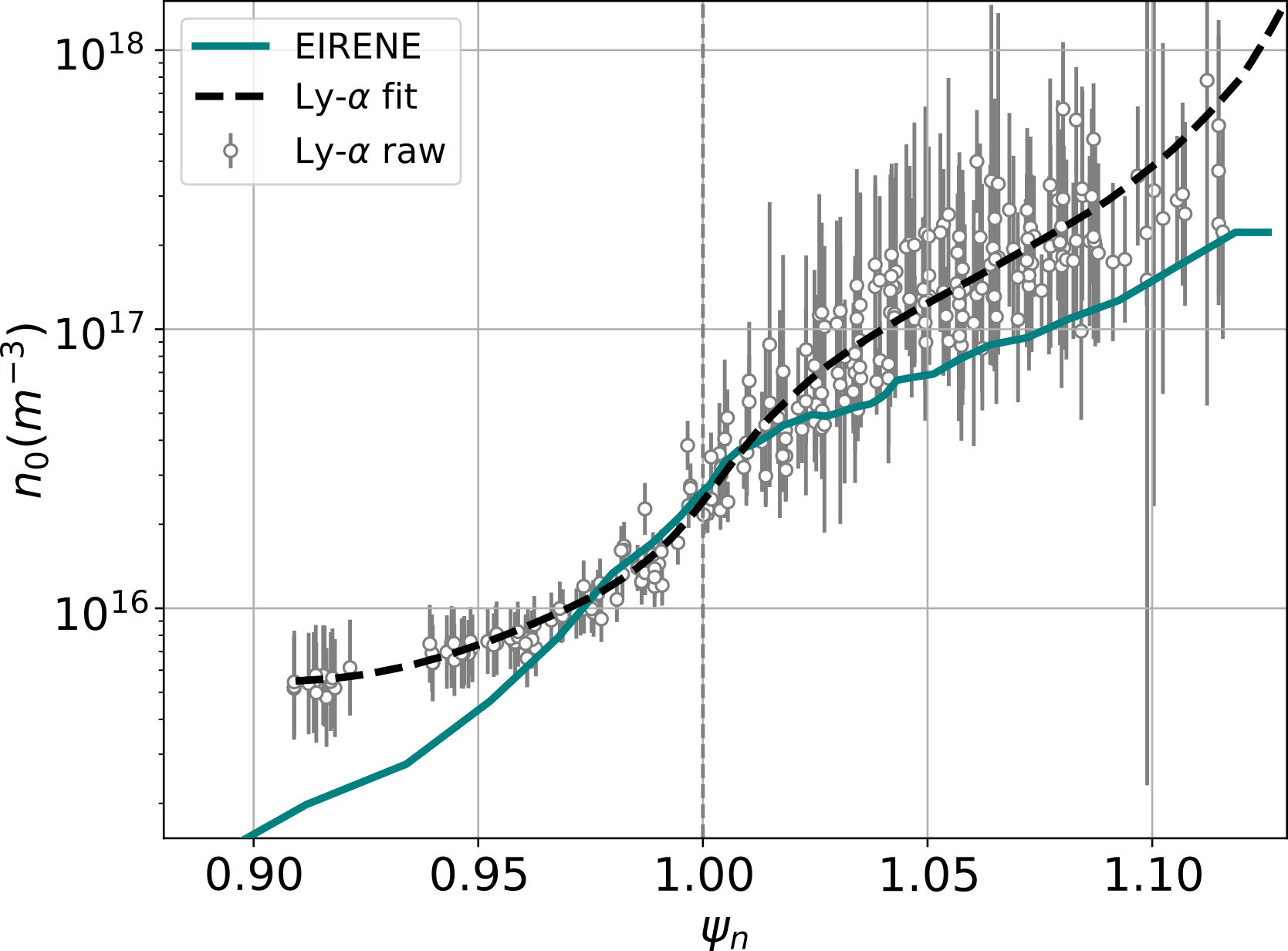}
\caption{The neutral density profile used to constrain the EIRENE solution, in log scale. In  open, gray circles are the points, along with the error bars, corresponding to the measured $n_{e}$ and $T_{e}$ from ETS, onto which $\epsilon_{\mathrm{Ly}_{\alpha}}$ is interpolated to experimentally infer $n_{0}$. The dashed black line represents the $n_{0}$ computed from the experimental profile fits. In turquoise is the simulated profile computed by EIRENE and interpolated onto the B2.5 plasma grid. This profile has been tuned using the procedure in Section \ref{subsec:neutrals_solps} to get as best a visual match in the pedestal region as possible.}
\label{fig:n0_solps_prof}
\end{figure}

\begin{table*}[hbtp!]
\begin{center}
\begin{tabular}{c|c|c|c|c|c}
 \hline
 Shot & $P_\mathrm{net}$ (MW) & $T_{e}^{95}$ (eV) & $n_{e}^{95}$ (10$^{20}$ m$^{-3}$) & $n_{0}^\mathrm{sep}$ (10$^{16}$ m$^{-3}$) & $\nu^{*}_\mathrm{sep} $\\
 \hline
 \hline
 1070821003 & 1.4 & 240 & 2.4 & 3.7 & 28 \\
 1070821004 & 2.0 & 280 & 2.5 & 3.0 & 24 \\
 1070821009 & 2.9 & 480 & 2.2 & 2.6 & 21 \\
 1070821008 & 3.4 & 700 & 2.3 & 2.5 & 13 \\
 \hline
\end{tabular}
\end{center}
\caption{Relevant parameters pertaining to the four selected discharges to be simulated using SOLPS-ITER. As before, the discharges at low $P_\mathrm{net}$ have high $n_{e}$, low $T_{e}$, and high $n_{0}$. The opposite is true for the discharges at high $P_\mathrm{net}$.}
\label{tab:solps_sim_params}
\end{table*}

\begin{figure*}
\centering
\includegraphics[width=1.9\columnwidth]{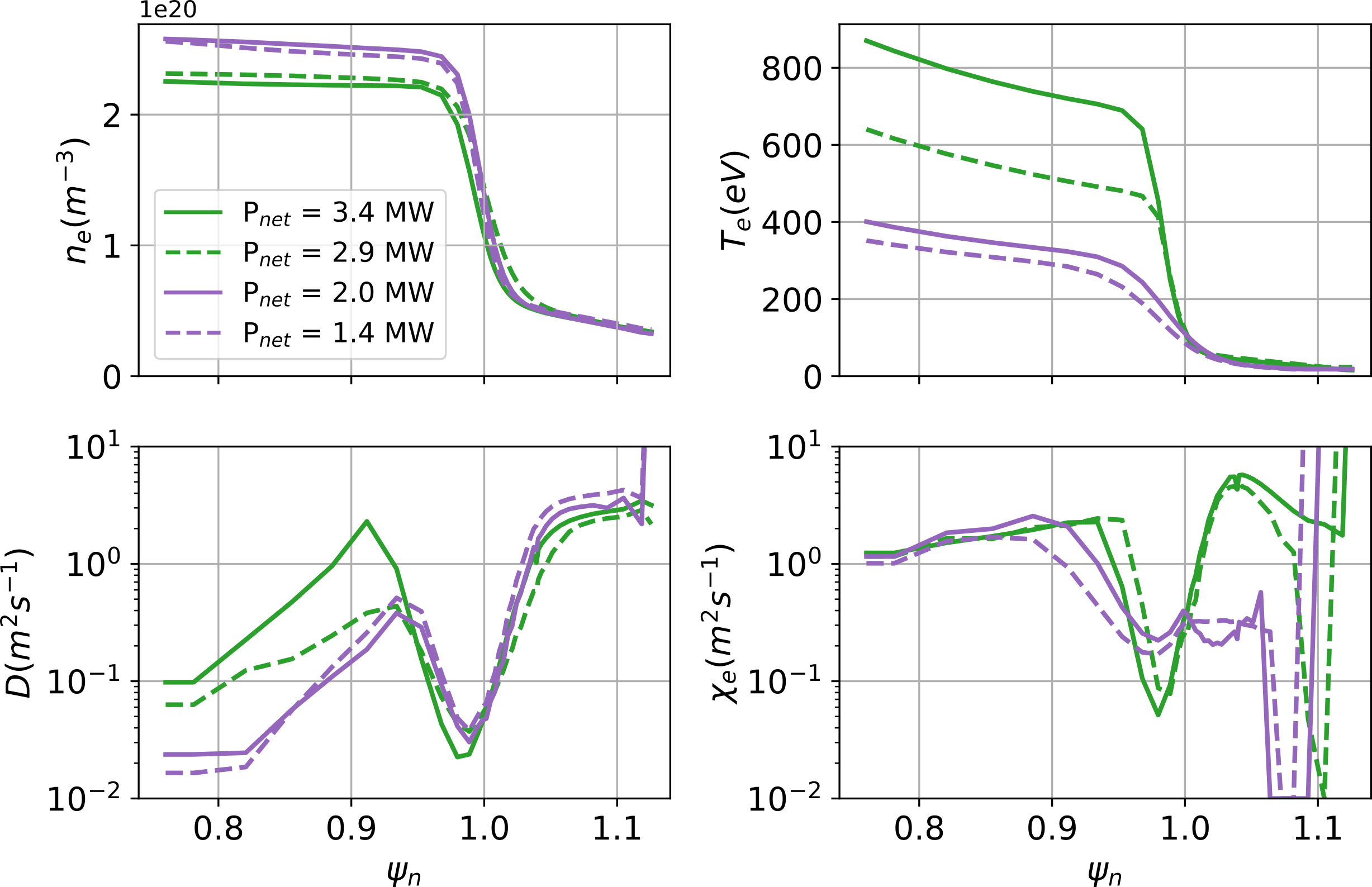}
\caption{Pedestal plasma solutions (top) and transport coefficient profiles (bottom) for the four simulated discharges tabulated in Table \ref{tab:solps_sim_params}. Quantities are shown for the particle transport channel (left) and for the electron thermal transport channel (right). $n_{e}$ and $D$ profiles are largely similar in shape. $T_{e}$ and $\chi_{e}$ profiles are not, varying significantly in height and in width. In the absence of $T_{i}$ measurements, $\chi_{i}$ is used simply to enforce $T_{e} = T_{i}$ and is not shown, as physical significance is uncertain.}
\label{fig:plasma_solns}
\end{figure*}

To determine transport profiles for the plasma in the main chamber and SOL, we employ an iterative approach used to reproduce TS profiles on NSTX and DIII-D \cite{canik_measurements_2011, wilcox_interpretive_nodate}. The approach uses a set of transport coefficient profiles at iteration $j$ to evaluate a new set of profiles at iteration $j + 1$ by using the experimental gradients, according to the following,

\begin{equation}
    D_{n}^{\mathrm{SOLPS},j+1} = -\frac{\Gamma^{\mathrm{SOLPS},j}}{\nabla n^\mathrm{exp}}
    \label{eq:dn}
\end{equation}

\begin{equation}
    \chi_{e,i}^{\mathrm{SOLPS}, j+1} = -\frac{q_{e,i}^{\mathrm{SOLPS},j} - \frac{5}{2} \Gamma^{\mathrm{SOLPS},j} T_{e,i}^{\mathrm{SOLPS},j}}{n^\mathrm{exp}\nabla T_{e,i}^\mathrm{exp}}
    \label{eq:chie,i}
\end{equation}

where $q_{e,i}$ is either the electron or ion heat flux. Equation \ref{eq:chie,i} is solved for electrons and ions separately. Since ion temperature, $T_{i}$, measurements are unavailable for these cases, we assume $T_{e} = T_{i}$, and set $\nabla T_{i}^\mathrm{exp} = \nabla T_{e}^\mathrm{exp}$ in the calculation of $\chi_{i}$. Earlier work comparing $T_{i}$ to $T_{e}$ showed that the temperature ratio, $\tau_{i} = T_{i}/T_{e}$, was not far from unity at the pedestal, but could be as high as six at the separatrix and in the SOL \cite{brunner_assessment_2013}. $\tau_{i}$ was found to be most clearly dependent on divertor collisionality - the more collisional the divertor, the more equilibrated $T_{i}$ and $T_{e}$. Given that the EDA plasmas in the current study were very collisional, $\tau_{i} = 1$, while likely an under-estimation, may not be wholly inaccurate, at least through the pedestal. Recent simulations of DIII-D have tested various approaches to setting the $T_{i}$ temperature. They show sensitivity of outputs, namely increased neutral pressure in the PFR at higher values of $T_{i}$ upstream. This motivates more detailed study of the sensitivity of upstream $n_{0}(\psi_{n})$, specifically in the pedestal, to choice of $T_{i}$ \cite{wilcox_interpretive_nodate}.

Figure \ref{fig:ne_dn_solps_profs} shows an example of experimental ETS data and the corresponding mtanh fit. Plotted also is the simulated $n_{e}$ profile as well as the corresponding $D_{n}$ profile used to match the experimental $\nabla n_{e}$ from the mtanh fit. Using this iterative transport solving scheme, between $25 - 35$\% of input power exits the grid radially across the outer grid boundary, presumably hitting the first wall and not arriving at the divertor. Furthermore, for simplicity, the code is run without including fluid drifts or currents. 

\subsection{Neutrals constraints}
\label{subsec:neutrals_solps}
A consequence of Equation \ref{eq:dn} is that without a constraint for $\Gamma^\mathrm{SOLPS}$, any number of values of $D_{n}$ can reproduce the requisite $\nabla n^\mathrm{exp}$. In other words, $n_{e}$ on its own is insufficient to constrain $\Gamma^\mathrm{SOLPS}$ and thus, $D_{n}$. As shown in Section \ref{sec:experiment}, knowing $n_{0}$ or equivalently $S_\mathrm{ion}$ fixes $\Gamma_{D}$, which does constrain the EIRENE calculation. Therefore, only if we can simultaneously match $n_{e}$, $T_{e}$, and $n_{0}$ at the OMP can we confidently make a conclusion about $D_{n}$, as well as $\chi_{e}$ \cite{reksoatmodjo_role_2021}, at least in a poloidally-averaged sense. 

Recent work on Alcator C-Mod has helped validate the $n_{0}$ calculation in EIRENE using the same LYMID camera \cite{reksoatmodjo_role_2021} across confinement modes. As mentioned in Section \ref{subsubsec:bcs}, rather than specifying a decay length given by $n_{e}$ and $D_{n}$ at the outermost grid cell, we use leakage BCs. These insist that the particle flux returning from the vessel walls be proportional to the local sound speed, $c_{s}$, and $n$, i.e. $\Gamma \sim \sqrt{\frac{T_{e} + T_{i}}{m_{i}}}n$. The current work finds that the proportionality constant for the particle channel, the particle leakage coefficient, $\alpha_{n}$, sets the magnitude of $\Gamma$ (and by consequence, $n_{0}$). This BC is thus used to shift $n_{0}$ up and down in magnitude, attempting primarily to match the magnitude of $n_{0}$ in the pedestal. While $\alpha_{n}$ changes the magnitude of $n_{0}$ it keeps the neutral gradient scale length, $L_{n_{0}}$, in the pedestal, $L_{n_{0}}^\mathrm{ped}$, i.e. the shape of the neutral density profile, relatively fixed. Previous studies of neutral penetration shows that $L_{n_{0}}$ depends strongly on $n_{e}$ \cite{mordijck_overview_2020}. Since we are seeking a match to $n_{e}$, we must also match $L_{n_{0}}$, and we instead mobilize the thermal leakage coefficients for electrons, $\alpha_{T_{e}}$, and ions, $\alpha_{T_{i}}$, to tune $L_{n_{0}}^\mathrm{ped}$. These essentially modify $q_{e,i}$, which indirectly affects $\Gamma_{D}$ by modifying the shape of $n_{0}$. The coefficients \{$\alpha_{n}$, $\alpha_{T_{e}}$, $\alpha_{T_{i}}$\} thus become the toolkit needed to match both the magnitude and slope of $n_{0}$ across the pedestal, given a particular B2.5 solution. Updating the EIRENE solution, however, also affects B2.5 by modifying the sources for the plasma transport equations. This essentially adds an outer loop to the iterative process described in Section \ref{subsec:transport_coefs}. Figure \ref{fig:n0_solps_prof} shows an example of the $n_{0}$ profile, calculated by EIRENE and interpolated onto the B2.5 grid, which results from this tuning process and which corresponds to the case shown in Figure \ref{fig:ne_dn_solps_profs}.

For these discharges, the simulation domain is treated as an isolated system, with no injection or removal of particles. Since Alcator C-Mod had molybdenum walls which did not readily absorb deuterium, we set the albedo to unity at all vessel walls, including targets. As mentioned in Section \ref{subsubsec:bcs} we impose $\Gamma = 0$ at the core boundary. Recalling that these H-modes were not actively puffed (see Section \ref{sec:experiment}), we also do not inject gas into the simulation domain via gas puffing. Particle balance is therefore achieved in these simulations simply by manipulating the initial particle inventory. 

\subsection{Transition in particle and thermal transport}

Given the computational cost of these simulations, we select four discharges from the power scan to model, covering a range in $P_\mathrm{net}$ and $\nu^{*}$. Two of these are below the critical $P_\mathrm{net}$ and two are above. Key parameters of the four simulated discharges are listed in Table \ref{tab:solps_sim_params}. Figure \ref{fig:plasma_solns} shows simulated $n_{e}, T_{e}, D_{n}$ and $\chi_{e}$ profiles across the pedestal and into the SOL for the four simulated discharges. The $n_{e}$ and $T_{e}$ profiles shown here all closely match the fitted experimental profiles, within fit error bars. As expected, the high $P_\mathrm{net}$ discharges have prominent $T_{e}$ pedestals and slightly lower $n_{e}$ pedestals. As $P_\mathrm{net}$ drops, $T_{e}^\mathrm{ped}$ plummets and $n_{e}^\mathrm{ped}$ grows to the saturated level seen earlier, in Figure \ref{fig:n_vs_Sion}. At this scale, it is hard to discern variation in separatrix conditions. At the pedestal top, however, it is evident that the green curves are at considerably lower $\nu^{*}$ than the purple curves, especially given the $T_{e}^{-2}$ dependence in $\nu^{*}$.

The bottom panels of Figure \ref{fig:plasma_solns} show the transport profiles found to reproduce $n_{e}, T_{e}$ \emph{and} $n_{0}$. In the pedestal, $D_{n}(\psi_{n})$ shifts down and slightly radially inwards only at the highest value of $P_\mathrm{net}$. There are no drastic changes to the structure of the profile in the steep-gradient region of the pedestal. This is not surprising given the apparent similarity in shape of the $n_{e}$ profiles and the small variation in profile gradients. $\chi_{e}$, on the other hand, clearly changes both in magnitude and in profile structure across the $P_\mathrm{net}$ scan. Indeed, the rise in transport observed in the particle channel is even larger in the electron thermal channel. The purple curves have much higher $\chi_{e}$ across the steep gradient region than the green curves. They also appear to shift slightly radially inward, in the opposite direction of $D_{n}$ at low $P_\mathrm{net}$. It follows then that the relative shift between $n_{e}$ and $T_{e}$ and weakening of $\alpha_\mathrm{MHD}$ mentioned in Section \ref{sec:os} could also be explained by the spatial decoupling of the particle and electron thermal transport shear layers in low $P_\mathrm{net}$ plasmas. More notable, however, is the significant widening of $\chi_{e}(\psi_{n})$ as $P_\mathrm{net}$ drops. The $T_{e}$ pedestals at low $P_\mathrm{net}$, shown in the top right panel, are considerably wider than those at high $P_\mathrm{net}$. At the separatrix, $D \sim 6 \times 10^{-2}$ m$^{2}$s$^{-1}$ and $\chi_{e} \sim 3 \times 10^{-1}$ m$^{2}$s$^{-1}$ for all values of $P_\mathrm{net}$. This change in $\chi_{e}$ best explains the weakening of the $\nabla T_{e}$ observed in Section \ref{sec:os}.

\begin{figure}[h!]
\centering
\includegraphics[width=0.9\columnwidth]{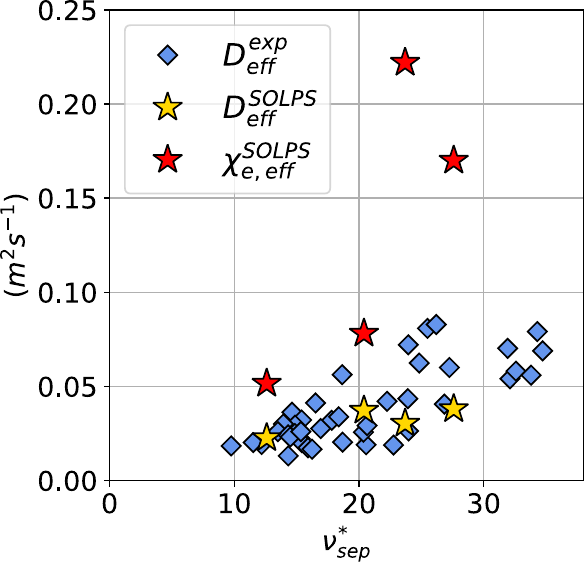}
\caption{Transport coefficients, inferred experimentally (diamonds) or computationally (stars) against separatrix collisionality. The blue diamonds are the experimentally inferred particle transport coefficients. The yellow stars are the same coefficients, but determined computationally using Equation \ref{eq:dn} iteratively. The red stars are the electron thermal transport coefficients determined via the simulation, using the electron component of Equation \ref{eq:chie,i}. $D_\mathrm{eff}$ coefficients are extracted at mid-$n_{e}$ pedestal, while $\chi_{e}$ is extracted at mid-$T_{e}$ pedestal. $D_\mathrm{eff}^\mathrm{SOLPS}$ fall within the scatter of $D_\mathrm{eff}^\mathrm{exp}$, while $\chi_{e}^\mathrm{SOLPS}$ is larger at high $\nu^{*}_\mathrm{sep}$ but closer in magnitude at low $\nu^{*}_\mathrm{sep}$.}
\label{fig:transport_nustar_solps}
\end{figure}

In Figure \ref{fig:transport_nustar_solps}, we plot the simulated $D_{n}$ and $\chi_{e}$ onto the experimentally inferred $D_\mathrm{eff}$, as a function of $\nu^{*}_\mathrm{sep}$. Indeed the simulated $D_\mathrm{eff}$ from SOLPS fall within the scatter of the experimental data, trending upwards with collisionality. This acts as a confirmation that the use of 1D analysis of plasma and neutral profiles as described in Section \ref{sec:experiment} yields a value for $D_\mathrm{eff}$ consistent with the poloidally-averaged one used in the SOLPS simulations. Of course, as different machines may have different poloidal neutral distributions, this confirmation is not in general the case across machines. Figure \ref{fig:transport_nustar_solps} shows that it is actually $\chi_{e}$ that undergoes the most dramatic transition at a critical collisionality, $\nu^{*}_\mathrm{sep,crit} \approx 22$. Interestingly enough, at $\nu^{*}_\mathrm{sep} < \nu^{*}_\mathrm{sep,crit}$, $D_{n} \sim \chi_{e}$. At $\nu^{*}_\mathrm{sep} > \nu^{*}_\mathrm{sep,crit}$, however, $D_{n} << \chi_{e}$. Recent computational work using pedestal gyrokinetics has indicated that the ratio of diffusivities in the pedestal in different transport channels may be indicative of different transport drives \cite{kotschenreuther_gyrokinetic_2019}. Indeed this may be consistent with the earlier assertion of transition in the character of turbulence from Figure \ref{fig:deff_alphat}.

\section{Discussion}
\label{sec:discussion}

As mentioned in Section \ref{sec:fueling}, pedestal stiffness and critical gradient-like behavior in the edge of Alcator C-Mod is not a new finding. Given C-Mod's proximity in parameter space to next-generation devices \cite{creely_overview_2020, vondracek_preliminary_2021, sorbom_arc_2015}, it is imperative to develop improved understanding for the conditions under which this phenomenon occurs. This paper proposes collisionally-driven transport as the mechanism by which a pedestal at its natural density responds to increased neutral flux and regulates its density gradient. As shown in Figure \ref{fig:n_vs_Sion}, as $S_\mathrm{ion}^\mathrm{sep}$ increases, so does $n_{e}^\mathrm{sep}$. Higher $S_\mathrm{ion}^\mathrm{sep}$ corresponds to an increase in $n_{0}^\mathrm{sep}$ as well, which presumably increases plasma thermal sinks, dropping $T_{e}^\mathrm{sep}$. The concomitant drop in $T_{e}^\mathrm{sep}$ ultimately drives pedestals that are dense with neutrals to higher collisionalities. Figures \ref{fig:deff_nu} and \ref{fig:deff_alphat} show that as the collisionality at the separatrix increases, so does $\alpha_{t}$, as well as $D_\mathrm{eff}$, likely as a result of a change in the nature of the transport. Figure \ref{fig:flux_grad} shows that when the gradient reaches a certain value, increased fueling increases transport, and at large enough values, the gradient begins to shrink. This additional resistive transport mechanism necessitates the use of varying diffusive transport coefficients. Given that these plasmas are all in the EDA regime, it is natural to question whether pedestal regulation via separatrix collisionality is inherently a feature of this regime. It has been shown on Alcator C-Mod that in EDA H-modes, the density fluctuations characteristic of the quasi-coherent mode (QCM) may be linked to increased particle transport \cite{terry_transport_2005}. Analysis to evaluate possible links between the QCM and the type of transport described here are ongoing.

It is important to note that this work says little about how a pedestal might reach a critical gradient to begin with and what types of devices might be susceptible to reaching one. Two characteristics of Alcator C-Mod, both hinted at above, appear important. The first of these is a neutral effect, which affects particle transport via indirect modification of collisionality. Having operated at the highest of plasma densities, C-Mod is proposed to be the device closest to the opaqueness predicted on both SPARC and ITER \cite{mordijck_overview_2020}. Inability to fuel inside the pedestal will push the ionization front radially outward, increasing collisionality at the separatrix or in the near SOL, thereby exacerbating the gradient regulation via resistive transport. This may very well occur at high densities in future machines. It is important to note that in terms of typical length scales, e.g. the poloidal gyroradius, the pedestal width, or the machine size, Alcator C-Mod is in a somewhat different parameter regime than future reactors. When considering dimensionless quantities, however, like opaqueness, or the gyroradius normalized to the minor radius, $\rho^{*}$, linked to both neoclassical and turbulent transport, this is not entirely the case. As for $\nu^{*}$, the parameter identified in this work as responsible for enhanced pedestal transport, large radial variation in this value makes it difficult to make conclusions about reactor-relevance, as noted in Section \ref{subsec:transport_pedestal}.

The second effect is related to the poloidal distribution of neutrals on C-Mod. It existed in the so-called main chamber recycling regime \cite{labombard_cross-field_2000}, consistent with large radial particle fluxes observed at the OMP. Measurements from a gas puff imaging system as well as from a stochastic model imply that this is linked to large filaments called ``blobs" propagating radially outward, depositing a large number of particles and heat to the main chamber walls \cite{garcia_intermittent_2013}. In either case, large particle fluxes to the MC walls means high main chamber neutral density, setting up a plasma environment conducive to the rise of RBMs at or near the separatrix. While it is unclear if the cause of high MC fueling on C-Mod is its high density or rather the close-fitting wall, these features will also be present on next-generation devices, and MC recycling may continue to comprise a large fraction of the fueling dynamics.

That these C-Mod pedestals exhibit what appears to be a limit to gradient growth has a number of implications for new fusion devices. To realize sufficient fusion gain, reactors will require high core densities. These must also be compatible with robust pedestals for good confinement. This work shows that good confinement can be lost when $P_\mathrm{net}$ drops below a critical value, $P_\mathrm{net}^\mathrm{crit}$, even while remaining in H-mode. For the set of discharges analyzed, this critical value corresponds almost exactly to L-H power threshold, $P_\mathrm{th}^\mathrm{L-H}$, calculated from the Martin scaling \cite{martin_power_2008}. This scaling has shown good agreement with power threshold experiments on C-Mod \cite{hughes_power_2011}. Across a wide range in $I_{p}$ and $\overline{n}_{e}$, some deviations from the scaling exist at low and high $I_{p}$ and $\overline{n}_{e}$, but for $I_{p}$ = 0.8 MA, the data matches the scaling \cite{ma_scaling_2012}. Since next-generation devices like SPARC and ITER, and later reactors, require high density and large plasma volumes, it is possible that they too will operate at $P_\mathrm{net}$ marginal to their corresponding $P_\mathrm{th}^\mathrm{L-H}$ \cite{hughes_power_2011, ma_scaling_2012, hughes_projections_2020}. 

It remains to be seen whether the observed transition to a collisional transport regime dictated by a collisional separatrix is inherently linked to $P_\mathrm{th}^\mathrm{L-H}$ in a quantitative way in general, or just by coincidence on C-Mod in this parameter range. Reactors, however, will likely require such a collisional separatrix with view of divertor handling \cite{Moulton_2021}, even if the pedestal is collisionless. Based on projections for $n_{e}^\mathrm{sep}$ and $T_{e}^\mathrm{sep}$ found in \cite{ballinger_simulation_2021, rodriguez-fernandez_core_2024}, $\nu^{*}_\mathrm{sep}$ on SPARC may range anywhere from below 1 and up to 10. Operation at even higher $n_{e}^\mathrm{sep}$ might bring this value closer to values observed here. Additionally, the discharges analyzed here do not include seeded impurities. The lack of impurities allows for an easier estimate of $Z_\mathrm{eff}$, which figures into $\nu^{*}$, facilitating the analysis in this work. This may differ, however, from a reactor scenario where impurities may be used to access divertor detachment. We also note that this paper only includes analysis of H-modes that are not actively gas puffed, even with D gas. While intrinsic wall-recycling may sound attractive, it means density control is substantially more challenging, or at least that there is a strong (although perhaps indirect) link between $P_\mathrm{net}$ and plasma fueling.

\section{Conclusions and future work}
\label{sec:conclusions}

This work presents new analysis of archival data from Alcator C-Mod. It deploys the routinely used ETS together with the less-frequently analyzed LYMID data to study the role of fueling and transport on pedestals in a high density, opaque plasma edge. Using inferences of ionization rates, it is observed that $n_{e}^\mathrm{sep}$ is a strong function of $S_\mathrm{ion}$, continuing to increase even at high $S_\mathrm{ion}$. $n_{e}^\mathrm{ped}$, on the other hand, stagnates, implying a saturated and even shrinking $\nabla n_{e}$. Comparing this gradient with $\Gamma_{D}$ shows a non-linear relationship, only explainable with varying $D_\mathrm{eff}$, despite relatively fixed shape and $q_{95}$. Inspection of profiles inside the pedestal top shows that the phase space at this radial location is best categorized by $p_{e}$ and $\nu^{*}$. Highly collisional pedestals are also found to have lower values of $\alpha_\mathrm{MHD}$, consistent with worsening H-mode quality and thus, $H_{98}$. It is found that $\nu^{*}_\mathrm{sep}$ is most influential in determining the large rise in $D_\mathrm{eff}$ across the pedestal and the subsequent flattening of the edge $n_{e}$ profile. Initial analysis links this to changes in $\alpha_{t}$, which mediates the transition betwen DW and RBM-dominated turbulence.

We supplement these experimental findings of changes to $D_\mathrm{eff}$ through modeling of selected discharges with the widely used edge code, SOLPS-ITER. High fidelity calculation of particle and heat sources, in addition to experimental 1D constraints, lends confidence to conclusions about how particle and heat transport must change to build up pedestal-gradients. Given the coupled nature of sources and transport demonstrated in this work, it recommends the development of models for the pedestal to include self-consistent accounting of both sources and transport. In particular, it allows for inference of $\chi_{e}$ (and $\chi_{i}$ to the degree that we trust the $\tau_{i} = 1$ assumption). Large growth in $\chi_{e}$ bolsters the claim that changes to the fundamental transport properties in these pedestals may be at play. Though not highlighted here, work has begun to understand how the poloidal distribution of neutrals may validate or limit the 1D inferences of $D_\mathrm{eff}$ from experiment. This, as well as understanding the contributions from different atomic processes to the neutral density in the pedestal, will be important steps in developing a model for the interplay between fueling and transport in the pedestal.

It has become evident that reactors will not be able to tolerate Type-I ELMs \cite{hughes_projections_2020, kuang_divertor_2020, leonard_impact_1999}. The non-ELMing behavior of the EDA H-mode, and other similarly non-ELMing modes, highlights the importance in understanding pedestal structure, and the mechanisms that determine it. While it is unclear if the EDA regime is achievable on reactors given that they will operate at high $T_{e}^\mathrm{ped}$ and thus, low $\nu^{*}_\mathrm{ped}$, some ELM-free solution is required. The analysis presented provides a useful experimental basis for how a pedestal may reach a transport limit before it is limited by less benign MHD events. It furthermore suggests that transport \emph{in} the pedestal can be closely linked to the collisionality at the separatrix, even more so than at the pedestal top. Even if $\nu^{*}_\mathrm{ped}$ were to remain low, as long as $\nu^{*}_\mathrm{sep}$ remains high enough, transport regulation of pedestal gradients may persist, ensuring ELM-free operation. Given that reactors may operate at high $n_{e}^\mathrm{sep}$ for divertor survivability, the requisite $\nu^{*}_\mathrm{sep}$ may not altogether be out of reach. As has been observed previously, high edge collisionality may also help widen near-SOL widths {\cite{eich_turbulence_2020, faitsch_broadening_2021}, helping work towards an integrated core-edge exhaust solution. 

\section*{References}

\bibliographystyle{unsrt}
\bibliography{references}

\end{document}